\documentclass[%
 reprint,
superscriptaddress,
 amsmath,amssymb,
 prc, floatfix,
]{revtex4-2}

\usepackage{graphicx}
\usepackage{dcolumn}
\usepackage{bm}

\newcommand{\dd}{\mathrm{d}}
\DeclareMathOperator{\sech}{sech}
\renewcommand{\Re}{{\mathrm{Re}}}

\usepackage{physics}
\newcommand{\braOket}[3]{\langle #1| #2 | #3 \rangle}
\newcommand{\braOketred}[3]{\langle #1\| #2 \| #3 \rangle}

\newcommand{\ii}{\mathrm{i}}
\newcommand{\etal}{\emph{et al.}}

\renewcommand{\vec}[1]{\bm{#1}}

\usepackage{color}

\usepackage{orcidlink}
\begin{document}

\title{Calculation of Dynamical Response Functions Using a Bound-state Method}

\author{Niels R. Walet\orcidlink{0000-0002-2061-5534}}
\email{Niels.Walet@manchester.ac.uk}
 \author{Jagjit Singh\orcidlink{0000-0002-3198-4829}}
\email{Jagjit.Singh@manchester.ac.uk}
\affiliation{%
 Department of Physics and Astronomy, University of Manchester, Manchester M13 9PL, UK}
 \author{Johannes Kirscher\orcidlink{ 0000-0002-9277-4999}}
\email{johannes.k@srmap.edu.in}
\affiliation{Department of Physics and Astronomy, University of Manchester, Manchester M13 9PL, UK}
 \affiliation{Center for Nuclear Studies, Department of Physics, The George Washington University, Washington DC 20052, USA}
 \affiliation{Department of Physics, SRM University-AP, 
Guntur District, Mangalagiri, Andhra Pradesh 522502, India}
 \author{Michael C.~Birse\orcidlink{0000-0003-4352-1730}}
\email{Michael.Birse@manchester.ac.uk}
\affiliation{Department of Physics and Astronomy, University of Manchester, Manchester M13 9PL, UK}
 \author{Harald W.~Grie{\ss}hammer\orcidlink{0000-0002-9953-6512}}
 \email{hgrie@gwu.edu}
 \affiliation{Center for Nuclear Studies, Department of Physics, The George Washington University, Washington DC 20052, USA}
 
 \author{Judith A.~McGovern\orcidlink{0000-0001-8364-1724}}
\email{Judith.McGovern@manchester.ac.uk}
\affiliation{%
 Department of Physics and Astronomy, University of Manchester, Manchester M13 9PL, UK}

\date{\today}

\begin{abstract}
We investigate a method to extract response functions (dynamical polarisabilities) directly from a bound-state approach applied to calculations of perturbation-induced reactions. 
The use of a square-integrable basis leads to a response in the form of a sum of $\delta$ functions. We integrate this over energy and fit a smooth function to the resulting stepwise-continuous one. Its derivative gives the final approximation to the physical response function. 
We show that the method reproduces analytical results where known, and analyse the details for a variety of models.  We apply it to some simple models, using the Stochastic Variational Method as the numerical method. Although we find that this approach, and other numerical techniques, have some difficulties with the threshold behavior in coupled-channel problems with multiple thresholds, its stochastic nature allows us to extract robust results even for such cases.
\end{abstract}

\maketitle

\section{Introduction}
Various methods have been developed to describe total and differential cross sections via square-integrable wave functions \cite{horiuchi_strength_2011,efros_method_2012,marcucci_hyperspherical_2020,drischler_toward_2021,zhang_fast_2022}. 
Another common approach in nuclear physics is to use the response functions (dynamical polarisabilities) in the calculation of cross sections \cite{walecka_theoretical_2004}.
As far as we are aware, the first method that combined these two ideas was the Lorentz-integral transform (LIT) \cite{leidemann_lorentz_2009,leidemann_modern_2013,leidemann_inversion_2020,marchisio_efficient_2003,efros_longitudinal_2004,efros_lorentz_2007,efros_method_2012,efros_calculating_2019,bampa_photon_2011,bacca_ab_2004,bacca_role_2009,bacca_electromagnetic_2014}. This uses a transformation of the response functions that leads to the solution of a set of inhomogeneous Schr\"odinger equations. A suitably weighted sum of overlaps between the solutions is then subjected to an inverse transform to obtain the response or dynamical polarisabilities. 

This procedure is reasonably robust for problems with a single open channel, such as the photo-dissociation of the deuteron \cite{bampa_photon_2011}, but seems more challenging in other cases (but see \cite{efros_calculating_2019,leidemann_inversion_2020,sobczyk_spectral_2022}).  Most standard nuclear-structure methods can be employed to generate a basis and, as we shall see below, good answers can be obtained with very different approaches in cases with a single open channel. For more than one open channel, however, all methods seem to struggle to some extent, something that we have found to be more obvious in those using the Stochastic Variational Method (SVM)~\cite{suzuki_stochastic_1998,mitroy_theory_2013}. 

Bound-state methods work with square-integrable basis functions, and any response calculated is naturally discrete. In the manner we apply this, this means that any response function becomes a sum of $\delta$  distributions with varying strengths, located at the eigenenergies of the channel Hamiltonian. In the LIT, this sum of $\delta$s gets folded with a finite-width Lorentzian \cite{leidemann_lorentz_2009}, giving a continuous function with a small but non-zero width. An inverse transform is then applied to this to get the physical response function. Typically, this inversion is done by fitting to the LIT of a continuous function constructed as a sum of carefully chosen basis functions \cite{leidemann_inversion_2020}. These basis functions should reflect properties expected of the physical, continuum response, such as the correct threshold behaviour. The choice of width also requires care: if too small, the discrete $\delta$s start to be resolved; but if too large, physical features can get washed out.  

Despite these subtleties, indications from the literature are that the method can work well for structured bases, such as hyperspherical harmonics (HH) or no-core-shell-model (NCSM) states  \cite{leidemann_lorentz_2009,leidemann_modern_2013,leidemann_inversion_2020,marchisio_efficient_2003,efros_longitudinal_2004,efros_lorentz_2007,efros_method_2012,efros_calculating_2019,bampa_photon_2011,bacca_ab_2004,bacca_role_2009,bacca_electromagnetic_2014}. This type of basis leads to a regularly-spaced spectrum of the kinetic energy operator. In contrast, the stochastically chosen basis of the SVM leads to a rather randomly-spaced spectrum, which can generate additional unphysical structures in the transformed response and so can be less well-suited for inversion of the LIT.

In this paper, we develop a powerful alternative which actually takes advantage of the randomness of the SVM. Instead of folding the response with a smearing function, this uses the \emph{integrated} response, that is, the integral up to some energy $\omega$ of a response function obtained with a square-integrable basis. This turns the sum of $\delta$ distributions into a step-wise continuous function of $\omega$. We then fit a continuous function to this, and differentiate to get an approximation to the physical response function. One could argue that this has just replaced one difficult problem (robust inversion of the LIT) to another (fitting of a function that is robust enough to yield a reliable derivative), but we will find that the latter is often an easier one. In this paper, we outline the method and present evidence that it is trustworthy, by comparing it with both an analytically solveable model and results from the LIT.  

The paper is organised as follows: in Sec.~\ref{sec:cx} we succinctly set out the underlying definitions; most of these can be found elsewhere in detail. We then, in Sec.~\ref{sec:model} show that for a simple P\"oschl-Teller potential, 
all the required quantities can be calculated in analytic form for the full continuum calculation, thus providing an important benchmark for basis-based methods. In the next section, Sec.~\ref{sec:L2}, we solve the problem using first a simple harmonic oscillator basis, and then 
a suitable Gau{\ss}ian basis. The latter will be shown to be extremely efficient. These are still relatively trivial problems, and we next compare to the photo-disintegration of the deuteron as discussed
by Bampa \etal~\cite{bampa_photon_2011}.  We see that both using the LIT transforms, and using a simple fit to an integrated response function, give largely identical
results, apart from small differences in areas of small response.

Next, we show that our approach well reproduces the response in a 3-particle continuum, Sec.~\ref{sec:threepart}, in anticipation of the difficulties we will encounter when we turn to a combination of a single particle continuum relative to a two-body bound state and a three-particle continuum in Sec.~\ref{sec:full}. We compare a hyperspherical-basis calculation and an SVM result for a two-channel model which has a two-body bound state in one of the channels.

Finally, we draw conclusions and give an outlook in \ref{sec:conc}.

\section{The calculation of the cross section\label{sec:cx}}
Inclusive cross sections due to an external probe (also called ``perturbation-induced") are typically of the form \cite{walecka_theoretical_2004}
\begin{equation}
    \frac{\dd^2\sigma}{\dd\epsilon \dd\Omega}
    =g^2\sum_i f_i(\epsilon,q,\Omega) F_i(\epsilon,\vec q)\,,
\end{equation}
where $\epsilon,\vec q$ with $q=|\vec{q}|$ are the energy and momentum transfer of the external probe to the system, $g$ is a coupling constant, $\Omega=\theta,\phi$ are the scattering angles, and the $f_i$ are kinematic factors.

The key aspect of  the calculation is thus the dynamical response functions \cite{marchisio_efficient_2003,efros_lorentz_2007}. For a single channel with no additional bound states, these  read
\begin{multline}
    F(\epsilon,\vec q)\\
    =\int \dd{\vec k} \braOket{\psi_0}{O^\dagger(\vec q)}{\psi_{\vec k}} \braOket{\psi_{\vec k}}{O(\vec q)}{\psi_0} \delta(E_{\vec k}-E_0-\epsilon)\,. \label{eq:F}
\end{multline}
The complete set of scattering eigenstates $\psi_{\vec k}$ are normalised as
\begin{equation}
   \int \dd{\vec k} \ket{\psi_{\vec k}}\bra{\psi_{\vec k}}=\widehat{1}\,.\label{eq:complete}
\end{equation}
As we shall see below, this can be calculated either directly, or by a Lorentz integral transform
\begin{equation}
    L(\sigma) =\int_0^\infty \dd \epsilon \frac{F(\epsilon,\vec q)}{(\epsilon-\sigma_R)^2+\sigma_I^2}\,
\end{equation}
where we take without loss of generality $\sigma_I>0$ but do not restrict $\sigma_R$.
Using completeness, Eq.~\eqref{eq:complete}, we can also express the Lorentz transform using the solution of an inhomogeneous Schr\"odinger equation, 
\begin{align}
    (H -E_0-\sigma)\ket{\tilde \psi(\sigma)} &= O(\vec q) \ket{\psi_0},\label{Eq:InSE}\\
    L&=\braket{\tilde \psi(\sigma)}{\tilde \psi(\sigma)}\,. \label{Eq:LITev}
\end{align}

In a finite basis of square-integrable functions, we can write the LIT decomposition in terms of the eigenfunctions $\phi_i$ and eigenvalues $E_i$ of the matrix representation of the Hamiltonian as discussed in \cite{efros_lorentz_2007}, i.e., 
\begin{equation}
    L(\sigma)=\sum_i \frac{|\gamma_i(\vec q)|^2}{(E_i-E_0-\sigma_R)^2+\sigma_I^2}\,,
\end{equation}
with \begin{equation*}
    \gamma_i(\vec q)=\braOket{\phi_i}{O(\vec q)}{\psi_0}\,.
\end{equation*}
Thus there is an explicit solution, which does not rely on the transform, as a sum of $\delta$ functions 
\begin{equation}
    F(\epsilon,\vec q)=\sum_i |\gamma_i(\vec q)|^2 \delta(\epsilon-E_i)\,,
\end{equation}
Clearly this form is not useful in itself, since we know that the solution obtained by using the scattering states is continuous. However, completeness shows that we can impose the the sum rule
\begin{equation}
    \sum_i {|\gamma_i(\vec q)|^2}=\braOket{\psi_0}{O^\dagger(\vec q)O(\vec q)}{\psi_0}\,,
\end{equation}
which is true for any complete basis, $L^2$ or not. This is useful to constrain the choice of basis functions.

Ref.~\cite{efros_lorentz_2007} suggests that increasing the number of basis functions leads to a denser spectrum, and thus a better LIT transformation; this is actually quite a subtle issue, as shown below. 
In the standard works on the LIT, one usually uses a nonzero $\sigma_I$ to smooth out small fluctuations when inverting the LIT. This removes detail from the response function, and we expect that
there is a price to pay in the inversion---at a minimum the displacement of reaction strength below the reaction threshold. That can be a small price if the response is smooth and phase-space factors suppress the cross section at low energies.
Nevertheless, we find it difficult to control the smoothness of numerical results, especially when using Gau{\ss}ian basis sets. In that case, it is much better to take some form of the limit $\sigma_I\rightarrow 0$, and here we will discuss one practical approach to approximate this limit. To this end, we
 shall mainly investigate the integrated response function 
 \begin{equation}
     T(\epsilon,\vec q)=\int_{0}^\epsilon\dd\epsilon' F( \epsilon',\vec q)\,,
 \end{equation} 
 rather than $F$.
 In the next section, we illustrate this approach for a simple one-dimensional potential.
    
\section{A simple model}\label{sec:model}
\subsection{Analytical treatment}\label{subsec:model_ana}
We shall use the one-dimensional P{\"o}schl-Teller potential \cite{poschl_bemerkungen_1933} as a test example (this potential is discussed in  some detail in the textbook Ref.~\cite{flugge_practical_1998}).
We consider the Hamiltonian
\begin{equation}
    H=-\frac{1}{2}\frac{\dd^2}{\dd x^2}-\frac{\lambda(\lambda+1)}{2\cosh^2x}\,.
\end{equation}
 The ground state is  $\psi_0(x)=(\cosh x)^{-\lambda}$ with eigenvalue $-\lambda^2/2$, and a first excited bound state $\psi_1(x)=\sinh(x)\psi_0(x)$ appears at
 energy $-(\lambda-1)^2/2$ for $\lambda>1$. 
Naively, we see there is second state at exactly zero energy for $\lambda=1$, which is one of the cases we will consider. This is not a problem since the potential is transparent and has no resonances in the continuum. However, it may be linked to some of the properties of the analytical solution below.
 
We now consider the single bound state case for $\lambda=1$,
which has a bound state energy $-1/2$ with normalised eigenfunction 
\begin{equation}
    \psi_0(x)=\frac{1}{\sqrt 2}\frac{1}{\cosh x}\,.
\end{equation}
We shall look at the ``photo-excitation" cross section for this model potential in the dipole approximation, $O(q)=\mathcal{E} q x$, and for simplicity we shall choose units where the product of electric field and charge $mathcal{E} q=1$. 

In order to evaluate Eq.~(\ref{eq:F}) directly, we note that  
the operator $x$ creates a state in the positive energy continuum. This means that we need the odd-parity positive-energy spectrum for energy $E=k^2/2$, which is known to be \cite{lekner_reflectionless_2007}
\begin{equation}
    \psi^\text{o}_k=\frac{1}{\sqrt{1+k^2}}\left(k \sin(kx)+\tanh(x) \cos(kx)\right)\,.
\end{equation}
We see that these wave functions approach $\sin(kx\pm\delta_k/2)$ for $x \rightarrow \pm \infty$, with the phase shift 
$\delta_k=2\arctan(1/k)$.

\subsubsection{Momentum-space basis}
From the result above, we calculate the generalised Fourier-decomposition of $x \psi_0$ relative to the energy eigenfunction $ \psi^\text{o}_k(x)$  in closed form, 
\begin{align}
    \tilde\phi(k)=\int_{-\infty}^\infty \psi^\text{o}_k(x)\, x\, \psi_0(x)\,\dd x=
    \frac{\pi  \sech \left(\pi k/2\right)}{\sqrt{2}\sqrt{k^2+1}}\,.\label{eq:psi1}
\end{align}
It is quite illustrative to calculate the inverse generalised Fourier transform of $\psi_1(k)$ in detail,
\begin{align}
  \phi(x):= \frac{1}{2\pi} \int_{-\infty}^\infty \psi^\text{o}_k(x) 
    \frac{\pi  \sech\left(\pi k/2\right)}{\sqrt{2}\sqrt{k^2+1}}\,\dd k\,.\label{eq:SE}
\end{align}
This integral can be  tackled by contour integration. Closing the contour in the lower half plane\footnote{actually, integrating on both sides of the negative imaginary axis until $z=-i$ as well},  we have contributions from the poles at $k=-(2n+1)\ii$, $n\in \mathbb{N} $. Taking the residues, we recover the original function as expected:
\begin{align}
    \phi(x)=&\frac{1}{\sqrt 2}    \biggl(
    \sinh (x)+x \sech(x)+\nonumber\\&
    \frac{1}{2}\sum _{n=1}^{\infty } (-1)^n (2n+1) \frac{(n-1)! }{(n+1)!}\sinh ((2 n+1) x)\nonumber-\\&
    \frac{1}{2}\tanh (x) \sum _{n=1}^{\infty } (-1)^n \frac{(n-1)!}{(n+1)!} \cosh ((2 n+1) x)\biggr)\nonumber\\
    =&\frac{1}{\sqrt 2} x \sech x=x\psi_0(x)\,.
    \end{align}

\subsubsection{Consistent normalisation}
We have not explicitly shown a consistent normalisation of the momentum space states; however since the integral
    \begin{equation}
         \int_{-\infty}^\infty x^2\sech^2 x\,\dd x=\frac{\pi^2}{6}\,.
    \end{equation}
   equals the corresponding  integral in $k$-space
    \begin{equation}
        \frac{1}{2\pi} \int _{-\infty }^{\infty }\frac{\pi^2 \sech^2\left(k \pi/2\right)}{\left(k^2+1\right)}\dd k=\frac{\pi^2}{6}\,,
    \end{equation}
    we see that the normalisation is consistent.

    \subsubsection{Response function}
       \begin{figure}
        \centering
        \includegraphics[width=\columnwidth]{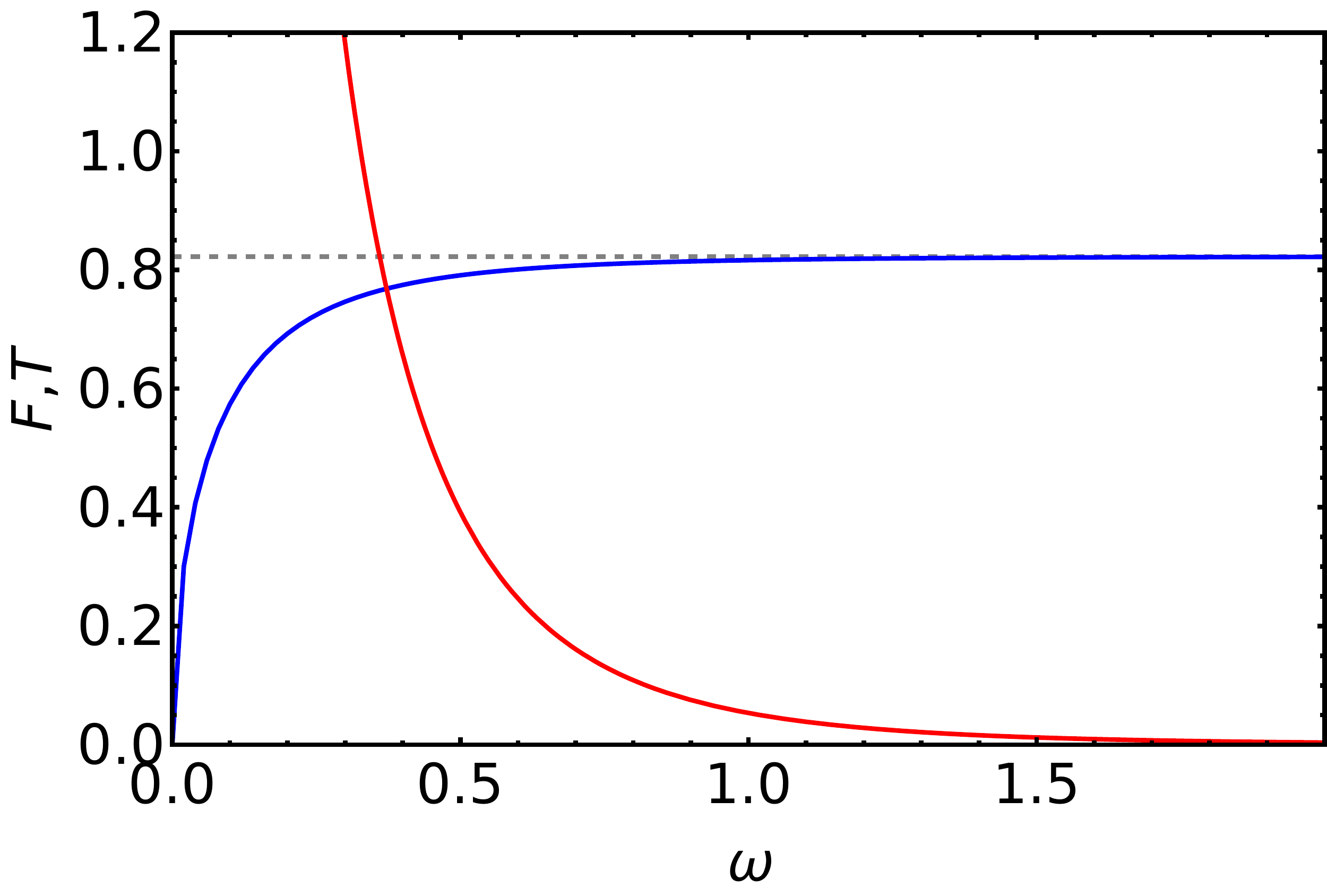}
        \caption{The analytic calculation for $F$, Eq.~\eqref{eq:Fex} (red curve), and the integrated response $T$, Eq.~\eqref{eq:Tex} (blue curve). The gray dashed line is the saturation value $\pi^2/12$.}
        \label{fig:baseline}
    \end{figure}
    
    The benchmark for all future calculations is the explicit analytic result for the response  function, which from Eq.~\eqref{eq:F} becomes
    \begin{equation}
        F(\omega)=\frac{\pi^2}{2} \frac{1}{\sqrt{2\omega}}
        \frac{\sech^2(\pi \sqrt{\omega/2})}{1+2\omega}\,,\label{eq:Fex}
    \end{equation}
    where $\omega=k^2/2$, which is the energy relative to threshold. This has an inverse square-root singularity at threshold, which is due to the constant density of states in one dimension.
    The integrated response distribution is then
    \begin{equation}
        T(\omega)=\int_0^\omega F(\omega')\dd \omega'\,,\label{eq:Tex}
    \end{equation}
    see Fig.~\ref{fig:baseline} for a representation of the integral and integrand. The most important message from that graph is the saturation of the integrated response function, which can be easily be proven to equal the norm  squared of $x\ket{\psi_0}$, 
    \begin{equation}
    \bra{\psi_0}x^2\ket{\psi_0}=\pi^2/12\approx 0.822467\,.
    \end{equation}

    To understand the analytical form of the LIT transformation, we express the response function as a $k$ space integral, 
    \begin{align}
    L(\sigma)&=\langle\psi_0|x (H-\sigma^*)^{-1} (H-\sigma)^{-1}x|\psi_0\rangle\nonumber\\
    &=
        \frac{1}{2\pi}\frac{\pi ^2}{2}\left[\int _{-\infty }^{\infty }\frac{ \sech^2\left(k \pi/2\right)}{\left(k^2+1\right) \left(\left(k^2/2-\sigma_R\right)^2+\sigma_I^2\right)}\dd k\right]\,,\label{eq:anaLIT}
    \end{align}
   where $\sigma=\sigma_R+\ii\sigma_I$. 
   
   \begin{figure*}
    \centering
    \includegraphics[width=\textwidth]{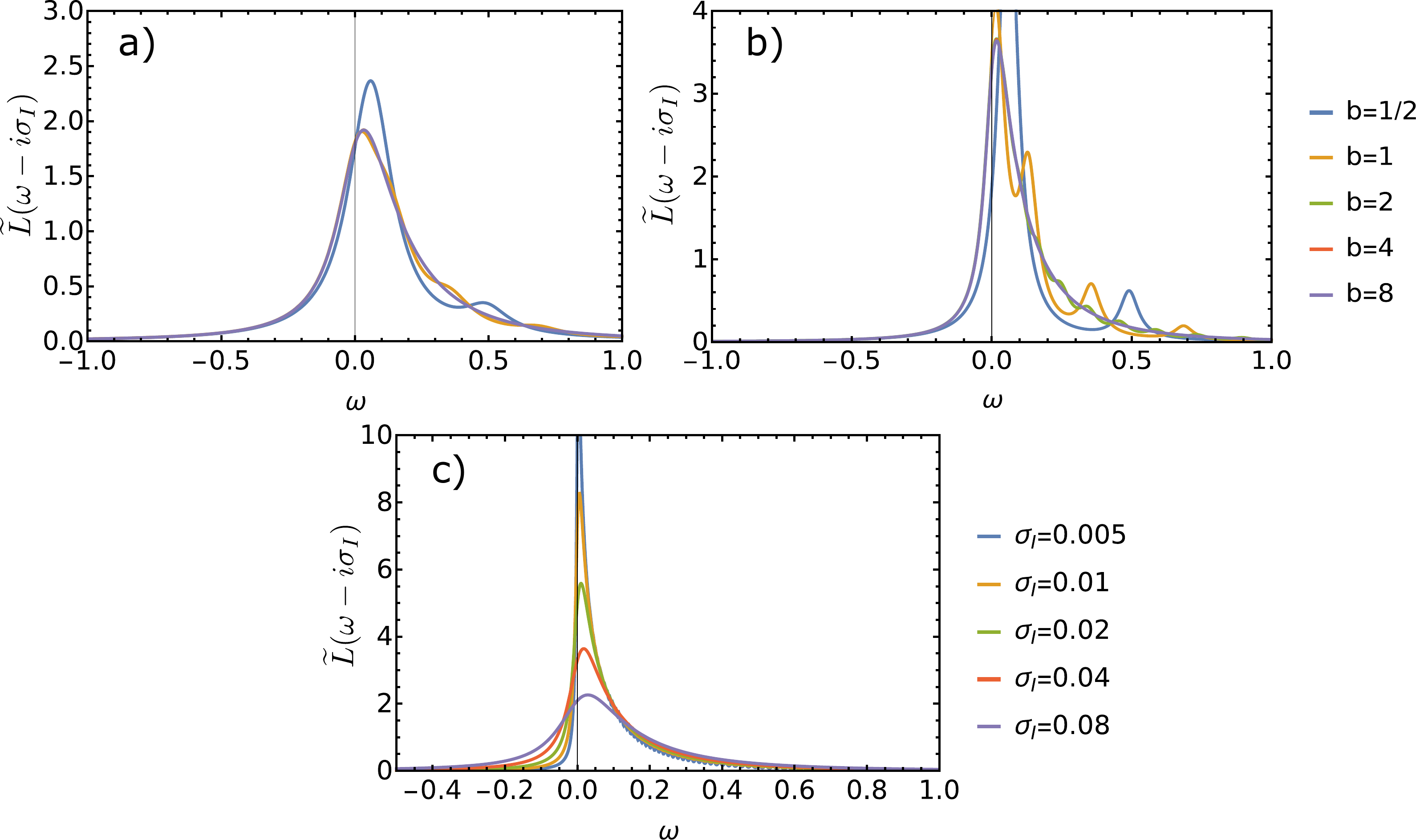}
    \caption{The normalised LIT transformation $\tilde{L}(\omega-i\sigma_I)$.
    a) and b) for fixed $\sigma_I$ ($0.1$ and $0.04$, respectively). We use various values of the harmonic oscillator length parameters, from coarse spectral spacing ($b=1/2$) to fine spectral spacing ($b=8$).  Case c) is for fixed $b=8$ (the closest spectral spacing) and various values of $\sigma_I$ as indicated. }
    \label{fig:LIT}
\end{figure*}

There are two classes of contributions to the integral, namely one from the pole of the Lorentzian and the other from the poles of the $\sech^2$, namely the poles of the wave function which lie in the lower half plane at $k=-(2n+1)\ii$ due to the specific contour chosen, see example above. The latter contribution is given by the sum
    \begin{widetext}
    \begin{align}
    L_1(\sigma)=&
\frac{6+\pi ^2}{12\left| \sigma +\frac{1}{2}\right| ^2}+
\frac{1}{\left| \sigma +\frac{1}{2}\right| ^4}+
\frac{16 \Re\left(\left(\sigma +\frac{1}{2}\right)^2\right)}{\left| \sigma +\frac{1}{2}\right| ^6}+
\frac{\Re\left(\frac{\psi ^{(1)}\left(\frac{1}{2}-\frac{\ii \sqrt{\sigma }}{\sqrt{2}}\right)-\psi ^{(1)}\left(\frac{\ii \sqrt{\sigma }}{\sqrt{2}}+\frac{1}{2}\right)}{\sqrt{\sigma } \left(\sigma +\frac{1}{2}\right) }\right)}{4 \sqrt{2}\sigma_I}
    \end{align}
    \end{widetext}
    where $\psi^{(n)}$ is the Polygamma function.  The shift of $\sigma$ by $1/2$ should be interpreted as the effect of the bound-state pole at $E=-1/2$.

    The poles of the Green's functions give a contribution to the integral in \eqref{eq:anaLIT} of
    \begin{align}
       L_2(\sigma)= \frac{\pi^2}{2 \sqrt{2} \sigma_{I}}
        \Re\biggl(\frac{\sech^2\left(\pi  \sqrt{\sigma/2}\right)}{\sqrt{\sigma} (2 \sigma+1)}\biggr),
    \end{align}
    which is singular in the limit $\sigma_I\rightarrow 0$.
    
    Both $L_1$ and $L_2$ contain a large contribution from the pole at $\sigma=-1/2$; these cancel substantially for small $\sigma$, and completely when $\sigma_I\rightarrow 0$. 
    Thus the response function can then be recovered from $L_2$; indeed
    \begin{equation}
        \lim_{\sigma_I\downarrow 0}\frac{\sigma_I}{\pi}L(\omega-\ii\sigma_I) =\frac{1}{\sqrt{2}\pi}
        \frac{\sech^2\left(\pi  \sqrt{\omega/2}\right)}{\sqrt{\omega} (2\omega+1)} \label{eq:LITsing}
    \end{equation}
    Moving into the complex $\sigma$ plane does therefore lead to a slightly confusing situation:
    The part of the response function that contributes to the result on the real axis is suppressed as we increase the imaginary part of $\sigma$, and a totally separate expression dominates for large $\sigma_I$. We see that the  dominant contributions are caused by the fact that the wave function has poles in the complex $k$ plane. The importance of the contribution of these poles of the wavefunction is in all likelihood specific to the P\"oschl-Teller potential, but the fact that the effect of the Green's function dominates on the real axis is probably generic and would explain why using an $L^2$ basis at small but nonzero $\sigma_I$ is normally effective in practical calculations.
    
\begin{figure}
    \centering
    \includegraphics[width=0.9\columnwidth]{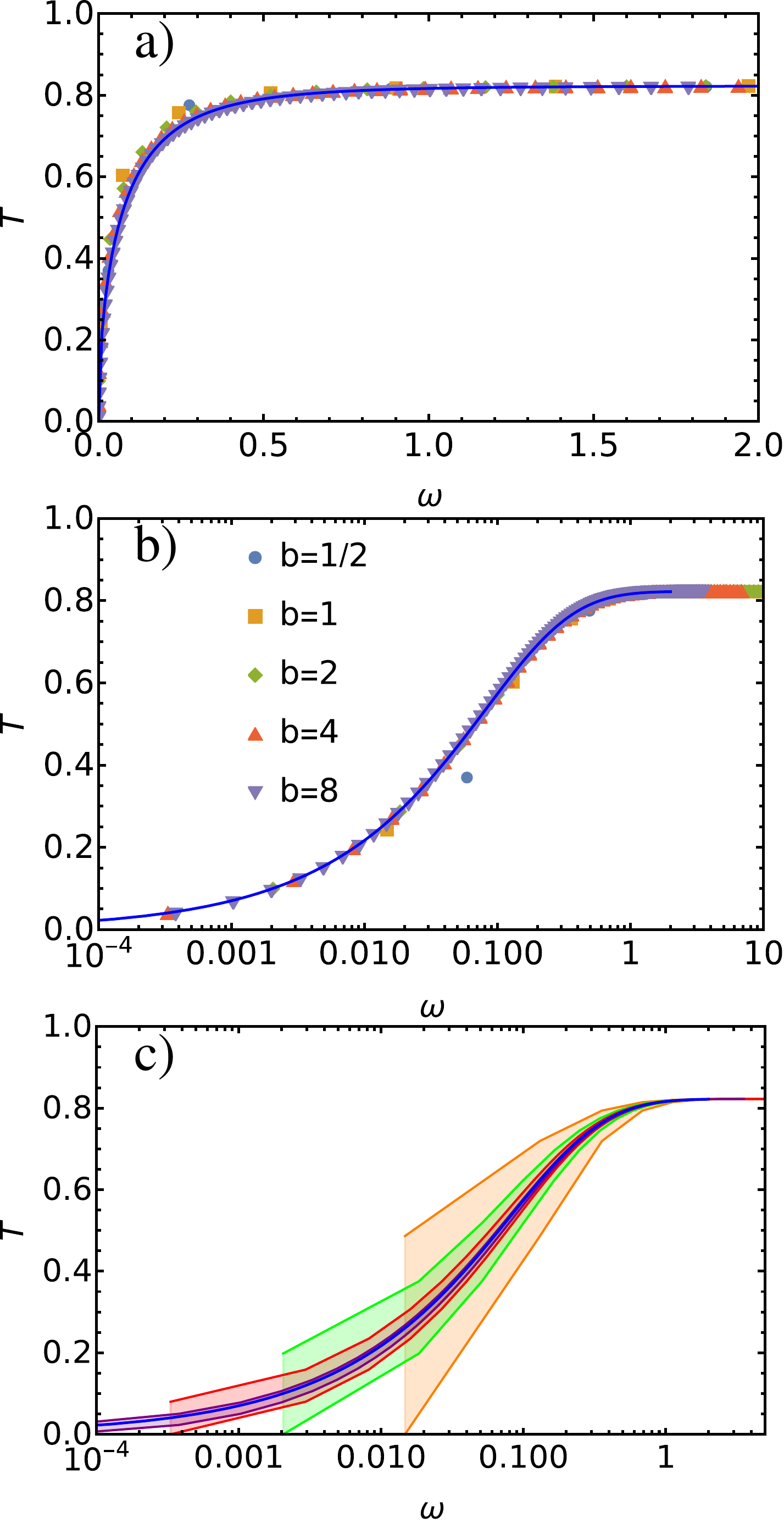}
    \caption{The cumulative response for various values of the the harmonic oscillator length parameter.
    In a) and b) the points plotted are the midpoints of each jump (see main text for details), the only difference between these two figures is the scale. The number of basis functions used is $b=1/2$: $60$, $b=1$: $50$, $b=2$: $80$, $b=4$: $120$ and $b=8$: $240$
    In each case, the solid blue curve is the analytical result. In c) the lower and upper lines connect the  values just before and just after the jumps, respectively. Colours correspond to a) and b) (we do not show $b=1/2$).}
    \label{fig:discrete}
\end{figure}
    
\subsection{$L^2$-basis calculations}\label{sec:L2}

In practical implementations of the methods discussed before, we normally use a square-integrable basis to find either the LIT transform, or the calculation of the integrated response function as advocated in this work.

We first look at the LIT transform where we  solve the equation \eqref{Eq:InSE}, where we now define $\sigma$ relative to the threshold at zero energy, 
\begin{equation}
    (H-\sigma )\tilde \psi(x)=x \psi_0(x)
\end{equation}
using a basis of $L^2$ functions. We shall use either a truncated complete basis consisting of a set of odd harmonic oscillator states, 
\begin{equation}
\chi^{(n)}(x)=H_{2n-1}(x/b)e^{-x^2/b^2}\,, \label{eq:hosc}
\end{equation} 
\begin{equation}
    \phi^{(n)}(x)=\frac{ (s \alpha^{n-1})^{3/4}}{\sqrt[4]{4\pi }}x \exp \left(-\frac{1}{2} s \alpha^{n-1}x^2 \right)\,,\label{eq:gaussb}
\end{equation} 
where a geometric progression with ratio $\alpha$ of widths starting at $s$ is used to ensure a numerically acceptable condition number of the overlap (norm) matrix. 

For the orthonormal basis \eqref{eq:hosc} we get
\begin{equation}
    \tilde\psi(x)=\sum_{n=1}^N c_n(\sigma) \chi^{(n)}(x),\quad d_n= \braOket{\chi^{(n)}}{x}{\psi_0}
\end{equation}
which gives the equation \begin{equation}
    (H_{nm}-\sigma\delta_{nm})c_m(\sigma)=d_n\,.\label{eq:discLIT1}
\end{equation}
If we denote the eigenvectors of $H$ in this basis as $\vec e_n$, and the as eigenvalues $\lambda_n$, we get a LIT transform of the form \eqref{Eq:LITev}, with $\gamma_i=\vec e_i \cdot \vec d$.

For the Gau{\ss}ian basis
\begin{equation}
    \tilde \psi(x)=\sum_{n=1}^N c_n(\sigma) \phi^{(n)}(x),\quad d_n= \braOket{\phi^{(n)}}{x}{\psi_0}\;,
\end{equation}
we define a Hamiltonian and overlap matrix by integration from the left with the same basis,
\begin{equation}
    \begin{Bmatrix}
    H_{nm}\\
    O_{nm}
    \end{Bmatrix}
    =
    \bra{\phi^{(n)}} 
    \begin{Bmatrix}
   H\\
    1
    \end{Bmatrix}\ket{\phi^{(m)}}
\,.\end{equation}
This gives the equation $(H_{nm}-\sigma O_{nm})c_m(\sigma)=d_n$.
If we denote the generalised eigenvectors of $H$ in this basis as $\vec e_k$ with eigenvalues $\lambda_k$, 
$H_{nm}(\vec e_k)_m=\lambda_k O_{nm}(\vec e_k)_m$, normalised as $(\vec e_l)_n O_{nm}(\vec e_k)_m=\delta_{lk}$,
we see that again $\gamma_i=\vec e_i \cdot \vec d$.

The integrated response thus always approaches a limit for large energies,
\begin{equation}
\sum_n (\vec e_n \cdot \vec d)^2\leq ||d||^2\,,
\end{equation}
where the last inequality is saturated for a complete decomposition of the state, e.g.,  $n\to\infty$.

Let us first analyse the LIT transformation, see Fig.~\ref{fig:LIT} for some representative results using a  harmonic oscillator basis. Since the Lorentzian $((x-\sigma_R)^2+\sigma_I^2)^{-1}$ is not normalised, the LIT transform diverges in the limit $\sigma_I=0$. Thus, for ease of comparison, it is better to look at the LIT transform of the normalised Lorentzian, $\widetilde L(\sigma)=L(\sigma)\sigma_I/\pi$.
We see that for a coarse spectral spacing (small $b$) we get a very oscillatory LIT transformation, from which it would be impossible to reconstruct the continuum result for the response function. For smaller spectral spacings we get a convergent result. However, the smaller the value of $\sigma_I$, the narrower the spacing we need in the spectrum to avoid spurious peaks and oscillations.  That raises the question how we can extract the inverse reliably from just one of these transforms. 
We thus need to answer two questions:
\begin{enumerate}
    \item How can one ensure a smooth inversion? We already know that the exact inversion of the data is given by a sum of delta distributions at the eigenenergies of the Hamiltonian \eqref{eq:discLIT1}, so for us to obtain a smooth result we can only perform an approximate inversion from a transform with substantial imaginary $\sigma_I$. 
    \item How do we disentangle the singular and regular parts of the LIT? We have seen above that we have two contributions, but only one survives in the limit $\sigma_I\rightarrow0$. At finite $\sigma_I$ both contribute, with domination from the ``wrong" component at positive energies.
\end{enumerate}
The second problem may be solved by doing the inversion at multiple values of $\sigma_I$, and interpolating to $\sigma_I=0$. This is a non-trivial task in realistic calculations, since oscillatory noise always enters the results. As explained before, some of this may be special for the P\"oschl-Teller potential--but if we have problems for one case, there may well be others.

The first one is more troublesome. In some sense what we are trying to do is to invert the LIT by sub-sampling, and solving the result by a continuous approximation, often by choosing a small set of basis functions that impose a shape on the resulting response function. Both of these are very indirect approaches, and it is almost unclear why we need the LIT as an intermediate if the inversion is such a difficult problem to resolve.

So, as an alternative approach, we consider the direct calculation of the response functions, given therefore simply as finite sum of delta distributions. It looks like we loose more than we gain. However, we now analyse the \emph{integral} of this sum as a function of the ``energy" $\omega=\sigma_R$ which is continuous, and can reliably be approximated by a smooth function.
This integrated response distribution $T$ is continuous and approximated by a smooth differentiable function, which can be used to reconstruct the response function $F$ itself.

We now analyse the results for the two basis sets. We have performed numerical calculations for harmonic oscillator bases with $b=1/2,1,2,4$ and $8$, see Fig.~\ref{fig:discrete}.
Since for the numerical calculations the integrated response $T$ makes a finite jump $(\vec d\cdot\vec e_n)^2$ at eigenvalue $\lambda_n$, there are multiple ways to represent this. We chose to plot at $\lambda_n$ the midway point between these two values. This seems to be overall a very true representation of the analytical result. As we can see from Fig.~\ref{fig:discrete}, we can closely mirror the analytic results. In c), we also show bands obtained from the outer and inner steps; we see a large width for a low spectral density, and a much narrower result for a high one: the width goes down rapidly as $b$ increases. Also, the midpoints track the analytical results remarkably well.

\begin{figure}
    \centering
    \includegraphics[width=\columnwidth]{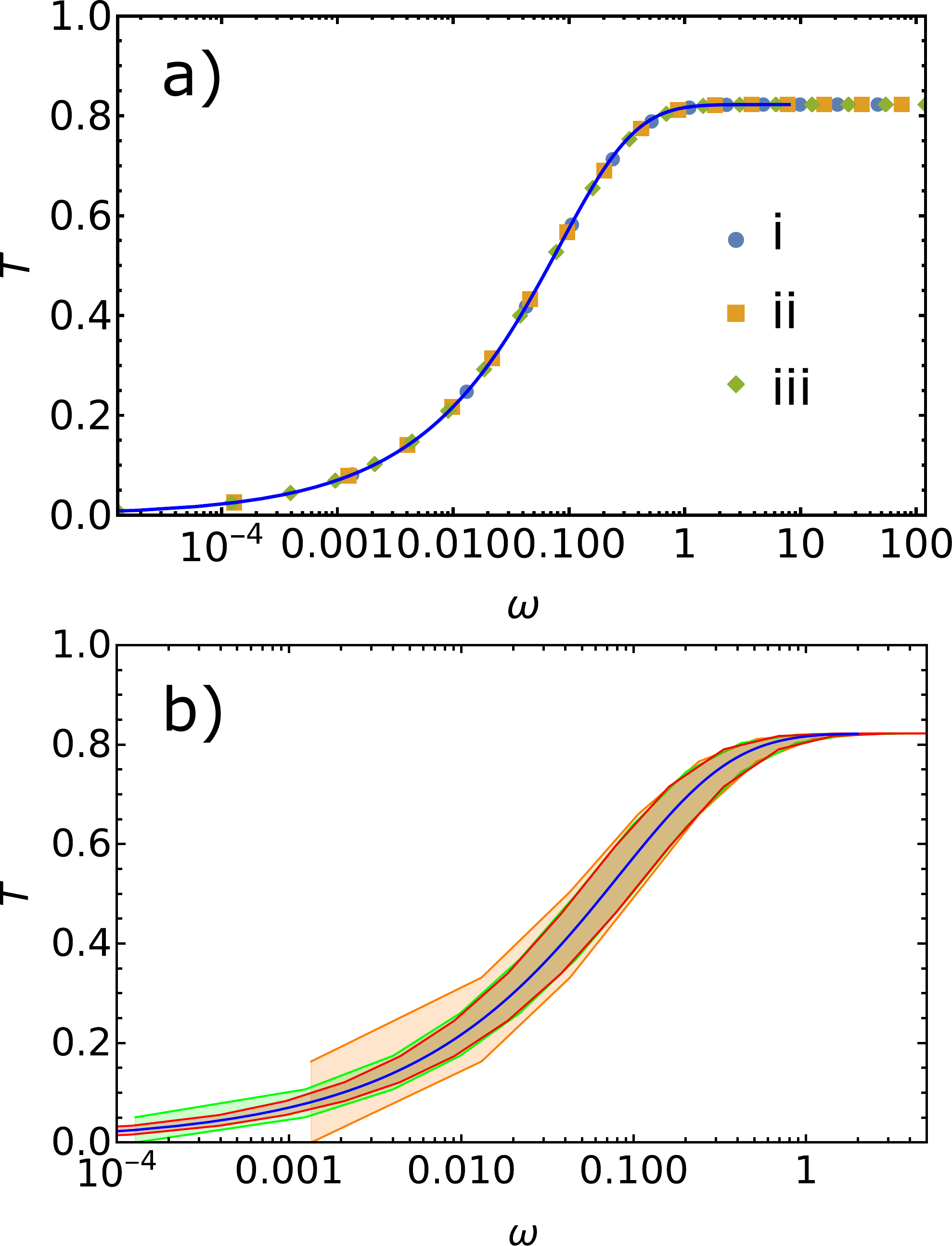}
    \caption{The convergence of the cumulative response for a small Gau{\ss}ian basis \eqref{eq:gaussb}. In a) we show the results for $\alpha=2$ and  $s=0.01$ (data set i, 12 basis functions),  $s=0.001$ (data set ii, 16 basis functions) and $s=0.0001$ (data set iii, 20 basis functions). As in Fig.~\ref{fig:discrete}, the points plotted are the midpoints of the jumps. In b) the lower
and upper lines of each colour connect the values just before and just after
the jumps, respectively. Colours correspond to those used in a).}
    \label{fig:discrete2}
\end{figure}

Since we want to yuse the SVM approach, which relies on Gau{\ss}ian basis functions, later on, we are particularly interested in the efficacy of using Gau{\ss}ian states as in Fig.~\ref{fig:discrete2}. We see that in all cases the midpoints of the jumps track the analytical results extremely closely. If we bracket the solutions with the upper and lower values of the jump, we see in the lower panel b) that the width is rather large. That is clearly an enormous over-estimate of the theory error
in the results, which are extremely close to the analytical results. We find that  $20$ Gau{\ss}ian basis states can achieve as much accuracy as $\approx200$ orthonormal ones.

\begin{figure}
    \centering
    \includegraphics[width=\columnwidth]{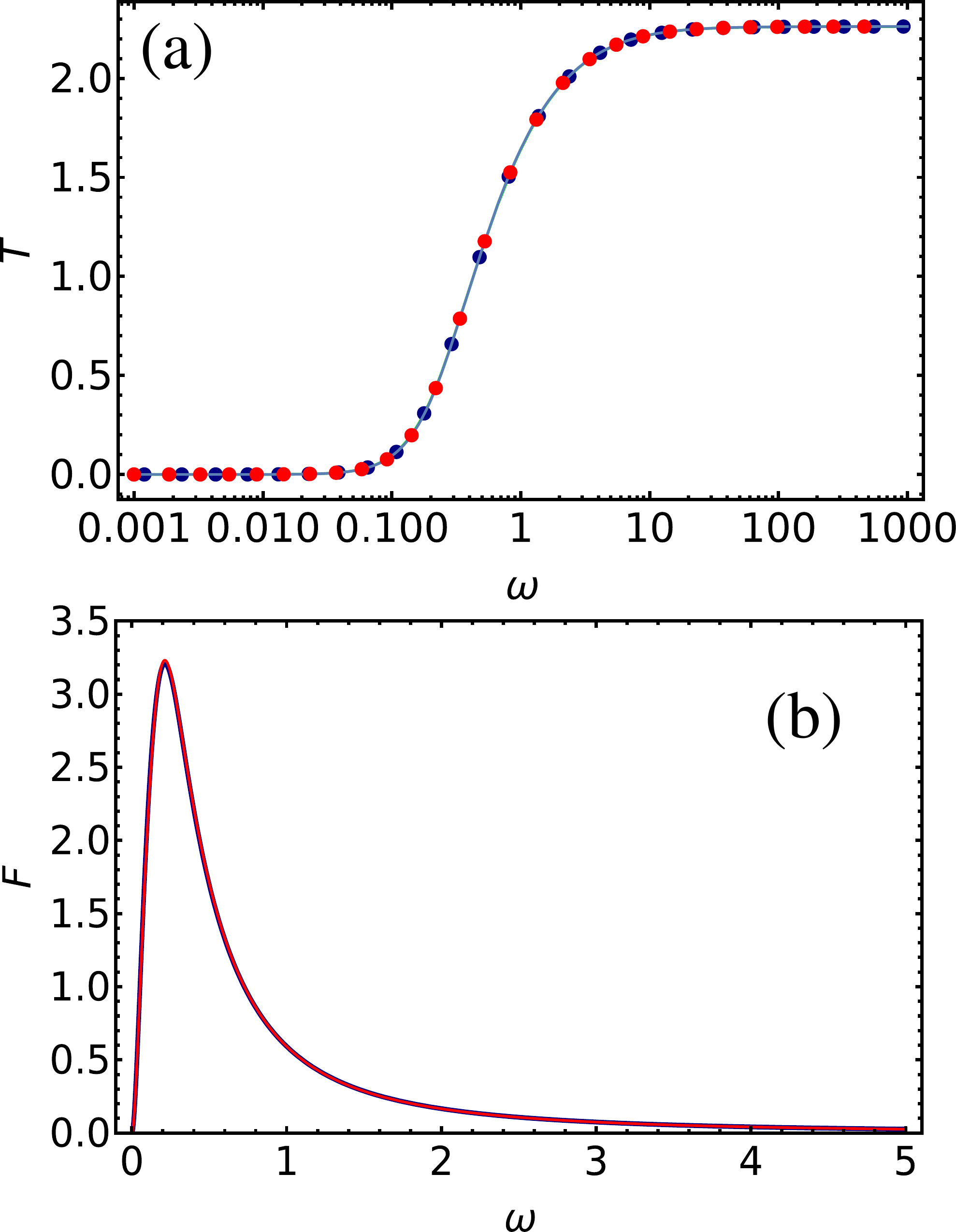}
    \caption{(a) calculation of the integrated response function (dots shown at midpoint of jump) for two different sets of Gau{\ss}ian basis functions (blue and red dots); solid lines are the fitted smooth response  distributions (b) response function obtained by differentiation of the fit from (a)--colours used correspond to the data sets. The two response distributions are visually indistinguishable.  
}    \label{fig:LIT3D}
\end{figure}


So far, we were concerned with a slightly artificial one-dimensional world. To show that the method set out above can be effective in a three-dimensional world and a much larger density of states above threshold, we consider the dipole excitation from an $s$-wave state, chosen as originating from a P\"{o}schl Teller potential,
\begin{equation}
    H= -\frac{1}{2r^2}\partial_r r^2 \partial_r -\frac{\lambda (\lambda+1)}{2} \frac{1}{\cosh^2r}\,,
\end{equation}
where we choose $\lambda=1.8$ for a single bound state at dimensionless energy $-0.32$ and  wave function
\begin{equation}
    \psi_0(r)=\frac{1}{r} \sinh(r) \cosh^{-\lambda}(r)\,.
\end{equation}
We now perform the calculation of the response function as set out above, with a perturbing dipole operator $z$. We diagonalise the $p$-wave Hamiltonian in a set of Gau{\ss}ian basis functions $z \exp(-\alpha r^2)$, calculate the integrated response function, and finally fit with a Pad\'e approximant of the form
\begin{equation}
    T(\omega)= \frac{\sum_{i=2}^{n} a_i \omega^i}{1+\sum_{i=1}^n b_i \omega^i}\,,
\end{equation}
where we take $n=4$ in the results shown in Fig.~\ref{fig:LIT3D}. This provides an almost perfect fit. 

We use two rather different basis sets but notice little difference in the results: the response functions, as shown in b) coincide within the line width. 

\section{Deuteron photo-disintegration}\label{sec:bampa}
As a further test of our method in a more physical context, we analysed the onebody part of the total $E1$ deuteron-photodisintegration cross section using the Argonne V18 potential\cite{wiringa_accurate_1995}. This exactly matches the treatment using the LIT  in Ref.~\cite{bampa_photon_2011}, allowing us to make direct comparisons with that work.

We use a numerical integration by mapping the interval $0\leq r<\infty$ onto  $x\in[0,1]$ via $r=\beta x/(1-x)$ and employ Gau{\ss}-Laguerre interpolation for a high-order approximation to the derivative, using the fact that we can continue the function as odd beyond $x=0$ and $x=1$.

\begin{figure}
    \centering
    \includegraphics[width=\columnwidth]{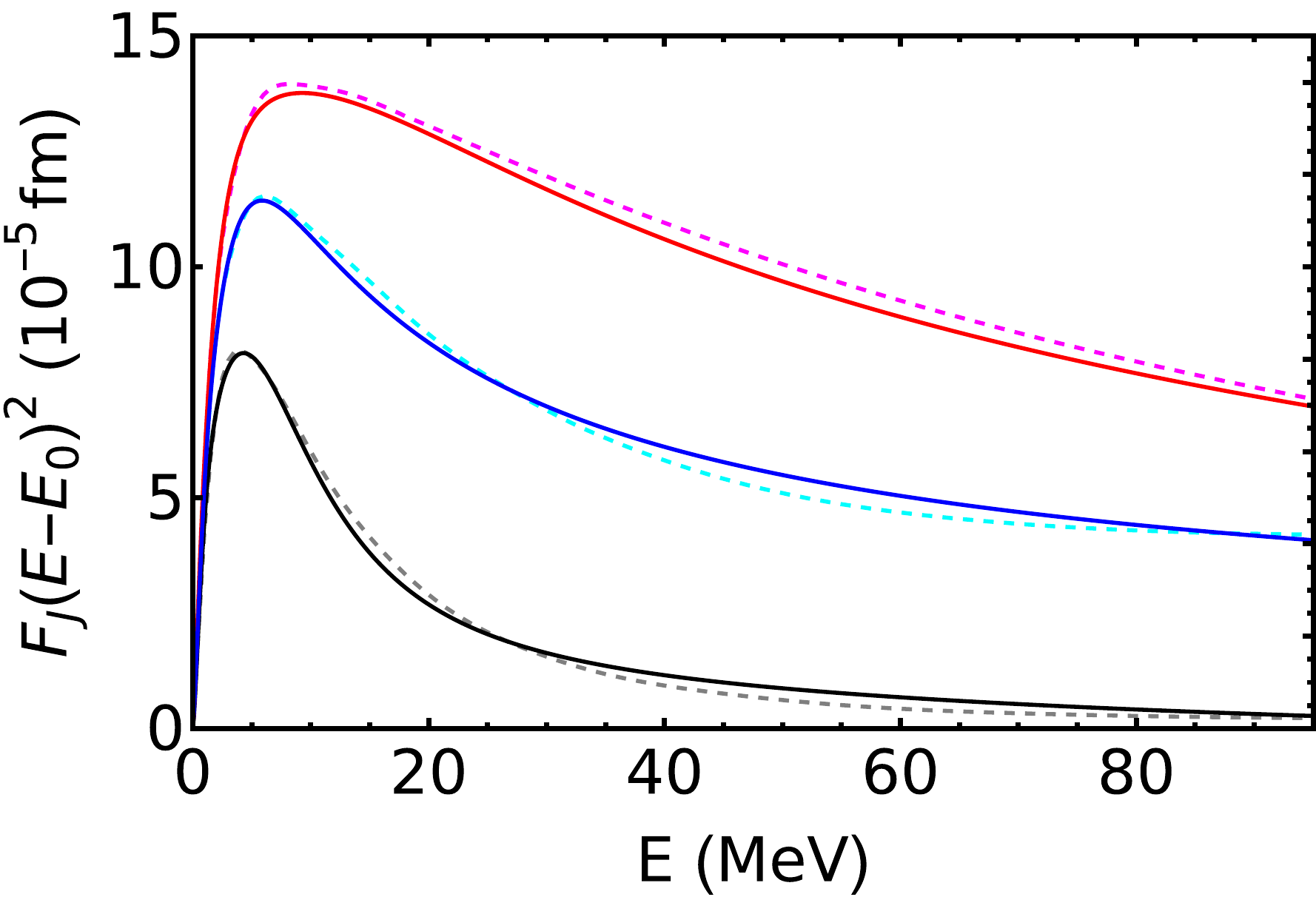}
    \caption{A comparison of the strength function multiplied with energy-transfer squared, Fig.~8 in Ref.~\cite{bampa_photon_2011}. Black line: $J=0$, blue line $J=1$ and red line $J=2$. The dashed lines are the results from Ref.~\cite{bampa_photon_2011}, solid lines are our results.}
    \label{fig:BampaFig8}
\end{figure}

As before, we calculate the cumulative response function $F$ by integrating over the $\delta$ functions at the eigenenergies of the intermediate channel Hamiltonian.
The resulting strength functions $F_J$ for intermediate-state total angular momentum $J$ are shown in Fig.~\ref{fig:BampaFig8}.  We see that our results very closely track the LIT ones from Ref.~\cite{bampa_photon_2011}. Small differences should come as no surprise, especially since the representation used in the figure enhances the small tails of these functions, where we expect the methods to show most the largest divergence. Bampa \etal~devote substantial work to the LIT inversion, and still see clear dependence on the basis function used. In our method, the fit is relatively robust, and again not as sensitive to the tail since it is very flat in the integrated function. Nevertheless, it is encouraging to see that the differences appear rather small.

\begin{figure*}
    \centering
    \includegraphics[width=1.5\columnwidth]{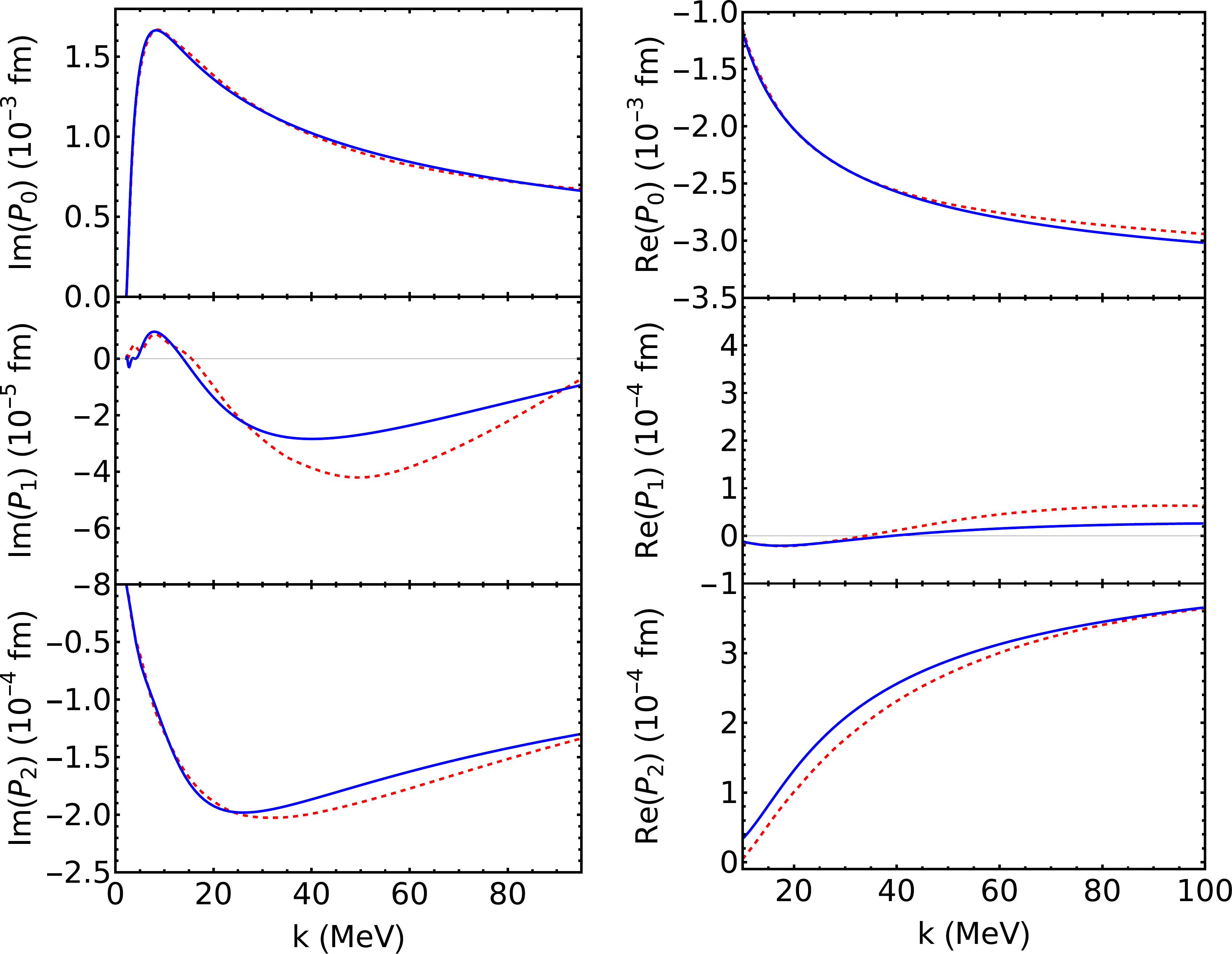}
    \caption{A comparison of the polarisabilities that enter the calculation of the cross section, Fig.~9 in Ref.~\cite{bampa_photon_2011}. Red  dashed lines show results from Ref.~\cite{bampa_photon_2011}, solid blue lines are results using the present method. 
    We have chosen the scale on the axes to be identical to that in the reference cited,
    and thus in the figures in the left column the $x$-axis runs from $0$ to $95$~MeV, and on the right from $10$ to $100$. }
    \label{fig:BampaFig9}
\end{figure*}

A more direct calculation of the basis of the cross section is given by the polarisation functions $P_J$. For a deuteron, the $E1$ operator allows for the channels $J=0,1,2$. The imaginary part is directly related to the $F_J$'s, and the real part can be calculated from a dispersion relation. This is a much more robust test of differences between the strength functions. Once again, we see in Fig.~\ref{fig:BampaFig9} that the results of the LIT and of our method are very similar. The main conclusion from the results on the left hand-side, which are direct calculation, is that there are small differences only--essentially, from top to bottom to middle we zoom in by a factor of 10 each time, and even at the largest scale the deviations are small.  Some of that seems to be due to a very robust calculation on our end, which explains the differences around $40$~MeV. The real surprise is how our simple fit seems to more easily capture the large-energy behaviour which required substantial work in the LIT calculation. That can also be seen in the right column. Since each of these curves is the result of a dispersion integral, it shows how close our results are to the LIT ones over a much larger domain. This gives us some confidence moving forward, even though we have one additional problem to discuss. 

\begin{figure}
    \centering
    \includegraphics[width=\columnwidth]{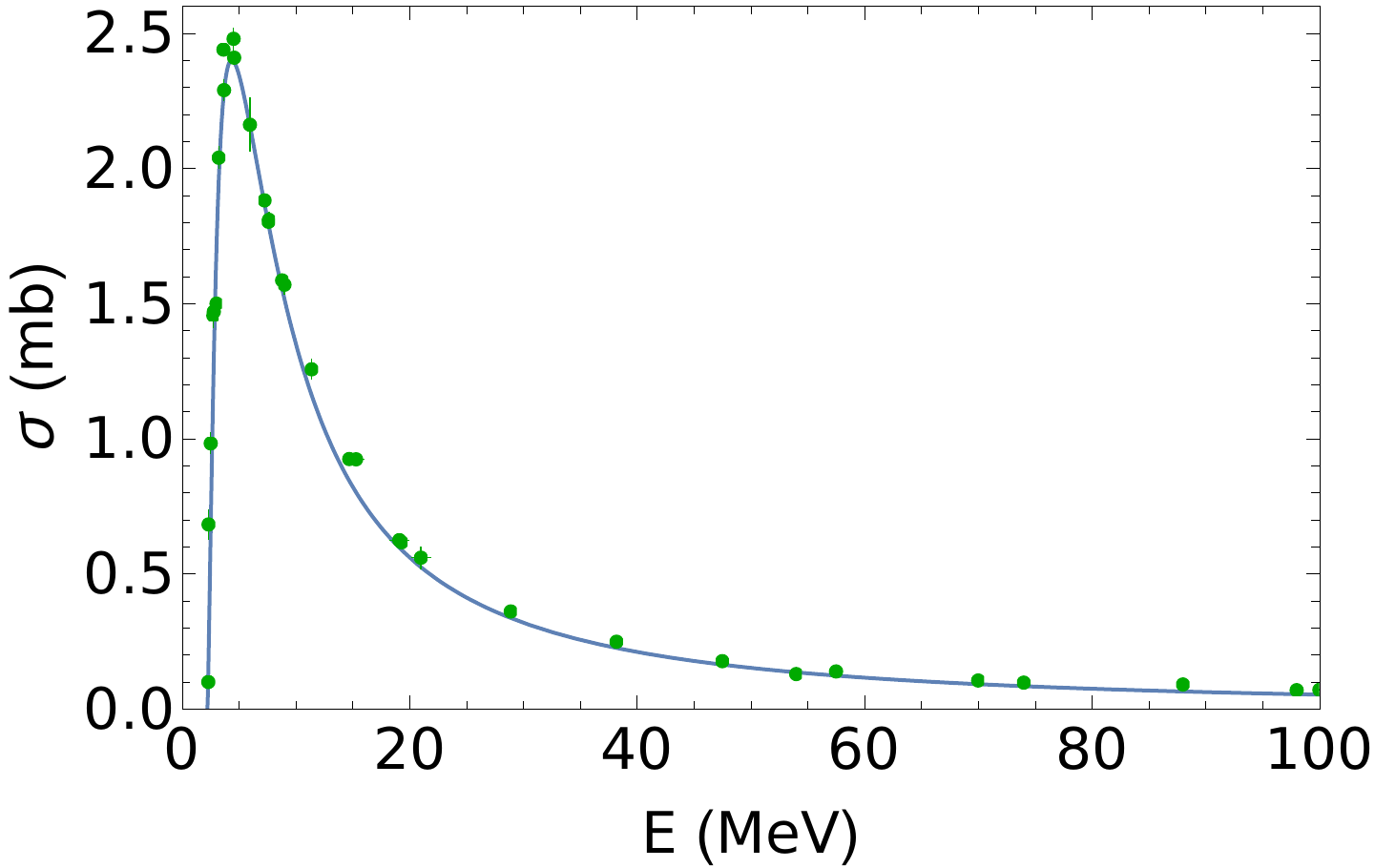}
    \caption{A comparison of the deuteron cross section as extracted from our method (using the imaginary part of the 
    responses in Fig.~\ref{fig:BampaFig9}) versus the data from the EXFOR database \cite{exfor}  (green dots). }
    \label{fig:Dxsec}
\end{figure}

We can use this output to calculate the differential photo-dissociation cross section, but little detailed data is available. Instead, we show in Fig.~\ref{fig:Dxsec} the total photo-disintegration cross section. The results from our calculation agrees well with the experimental data extracted from EXFOR \cite{exfor}. As we can see, the agreement is excellent, which is a confirmation both of the stability of the calculation and of the fact that the differences to the LIT variant are small. There is some indication that our results decay slightly too fast at the highest energies. That might not be surprising since our computation only contains the one-body $E1$ operator. While the importance of magnetic multipoles is well-known to decrease with energy, higher electric multipoles and pion-exchange currents play a more prominent at higher energies.

\section{Three-particle decays}\label{sec:threepart}
Clearly the method set out above is very effective for simple structureless two-body problems. One of the questions we have to answer is how it survives in the many-body context, especially where we may have multiple thresholds that complicate the calculations and their interpretation--indeed, this is one of the problems that led to the current approach.
The issue already arises for $A=3$, where both two-body and three-body breakup channels with different thresholds exist, e.g. $^3$He$\to ppn$ or $\to pd$.
One issue is the degeneracy of the three-body continuum: Since it is described by two Jacobi coordinates,
we expect a substantial degeneracy as a function of the excitation energy, which is apt to lead to some complications. 

So let us again study a simplified, exactly solvable, problem which, as we shall show later, is an illuminating simplification of the problem to be studied in detail in future articles. We consider the dipole response function
for a simple three-body ground state consisting of two identical bosons interacting with a third distinguishable particle, all three of the same mass. The dipole operator either acts on the third particle, or equivalently on the two bosons: translation invariance shows these two are identical up to a multiplicative constant. This is obviously the bosonic equivalent to the $^3$He system mentioned above.

To find a problem that can both be tackled analytically and numerically we start from the wave function
\begin{align}
    \psi_0&=\frac{1}{4\pi}\frac{4\alpha^{3/2}}{\sqrt{\pi}}\exp(-\alpha/2(r_{13}^2+r_{23}^2))\nonumber\\
    &=\frac{1}{4\pi}\frac{4\alpha^{3/2}}{\sqrt{\pi}}\exp\left(-\alpha \eta_3^2-\alpha \xi_3^2/4\right),
\end{align}
(see \eqref{Eq:Jacobi} for the definition of coordinates) and act with the dipole operator
\begin{equation}
    d_z=(Z-z_3)=\frac{1}{3}(z_{13}+z_{23})=\frac{2}{3}\eta_{3z}.
\end{equation}

We assume that the Hamiltonian describing states in the $L=1$ channel is just the free Hamiltonian, which takes the form 
\begin{equation}
    H^{(1)}=\frac{1}{2}(p_1^2+p_2^2+p_3^2)=\frac{1}{6} p_R^2+\frac{3}{4}p_\eta^2+p_\xi^2\,.
\end{equation}
This can be tackled with results discussed in Appendix \ref{app:model}.
We can evaluate the effect
of the $E1$ operator using only the ``12-3" Jacobi coordinates. 
We once again apply Eq.~\eqref{Eq:InSE}. Looking at the right hand side, we see the effect of the $E1$ operator on $\ket{\psi_0}$ gives an $L=0$ solution in $\xi$, and a $L=1$ state in $\eta$,
\begin{align}
    O_{E1}\ket{\psi_0}=\frac{1}{4\pi}\frac{8\alpha^{3/2}}{3\sqrt{\pi}} 
    \eta_{3z}\exp\left(-\alpha \eta_3^2\right)
    \exp\left(-\alpha \xi_3^2/4\right)\,.
\end{align}
The total response is the norm of this state,
\begin{equation}
    \braOket{\psi_0}{O_{E1}^2}{\psi_0}=\frac{1}{9\alpha}\,.
\end{equation}

In order to find the response function, 
we need to find the relevant (normalised) continuum solutions, which are
\begin{align}
    f_0(q_\xi,\xi_3)&=\frac{1}{4\pi}\sqrt{\frac{2}{\pi}}j_0(q_\xi \xi_3)\nonumber,\\
    f_1(\vec{q}_\xi,\vec \xi_3)&=\frac{3}{4\pi}\sqrt{\frac{2}{\pi}}\cos\theta_\eta \cos\theta_{q_\eta} j_1(q_\eta \eta_3)\,.
\end{align} The energy for the direct product of these two states is
\begin{equation}
    E=\frac{3}{4} q_\eta^2+q_\xi^2\,.\label{eq:Efree}
\end{equation}
\begin{widetext}
The normalisation can be checked from: 
\begin{equation}
   (4\pi)^2 \int_0^\infty q^2\,\dd q\, f_0(q,\xi) \int_0^\infty {\xi'}^2 \dd \xi' f_0(q,\xi') \exp(-\beta {\xi'}^2)=\exp(-\beta {\xi}^2)
\end{equation}
and
\begin{equation}
   (2\pi)^2\int_0^\infty q^2\dd q\,\dd\cos\theta_q f_1(\vec{q},\vec \xi)\int_0^\infty {\xi'}^2 \dd \xi'  d\cos\theta'
   \,f_1(\vec{q},\vec \xi') \xi'\cos\theta' \exp(-\beta {\xi'}^2)=\cos\theta\, \xi \exp(-\beta {\xi}^2)\,.
\end{equation}
Thus
\begin{align}
    \braOket{\vec q_\xi,\vec q_\eta}{d_z}{\psi_0} &=q_\eta  \cos (\theta_{q_\eta}) \exp\left({-\frac{q_\eta^2}{4 \alpha }-\frac{q_\xi^2}{\alpha }}\right)\frac{1}{3 \pi ^{3/2} \alpha ^{5/2}}\,.
\end{align}
This satisfies the consistency check
\begin{equation}
    (4\pi)(2\pi) \int \braOket{\vec q_\xi,\vec q_\eta}{d_z}{\psi_0}^2 q_\eta^2\dd q_\eta \dd\sin\theta_{q_\eta} q_\xi^2 dq_\xi = \frac{1}{9\alpha}\,.
\end{equation}
If we now substitute $q_\xi^2=\omega-\frac{3}{4}q_\eta^2$, and integrate over $\vec q_\eta$, we get, using $q_\xi=\sqrt{\omega-3q_\eta^2/4}$,
\begin{align}
    F(\omega)&=8\pi^2\int_0^{\sqrt{4\omega/3}} q_\eta^2 \dd q_\eta\int \dd\sin\theta_{q_\eta}
   \braOket{\vec q_\xi,\vec q_\eta}{d_z}{\psi_0}^2\frac{1}{2 q_\xi} \nonumber\\
&=
\frac{1 }{81 \sqrt{3} \alpha ^2} e^{-4 \omega /(3 \alpha )}\left(\left(\frac{\omega}{\alpha}\right)^2  I_0\left(\frac{2 \omega }{3 \alpha }\right)+\left(\left(\frac{\omega}{\alpha}\right)^2  -3 \frac{\omega}{\alpha}  \right) I_1\left(\frac{2 \omega }{3 \alpha }\right)\right)\,.\label{eq:I0I1}
\end{align}
\end{widetext}
Indeed, this still integrates to the same value as before.
The calculation for the integrated response can easily be done numerically, and we conclude that the function $F$ rises as $\omega^3$ for small $\omega$, and thus with a quartic power for the integrated response, and decays for large $\omega$ as $\exp(-2\omega/3\alpha) \omega^3/2$.

\begin{figure}
    \centering
    \includegraphics[width=\columnwidth]{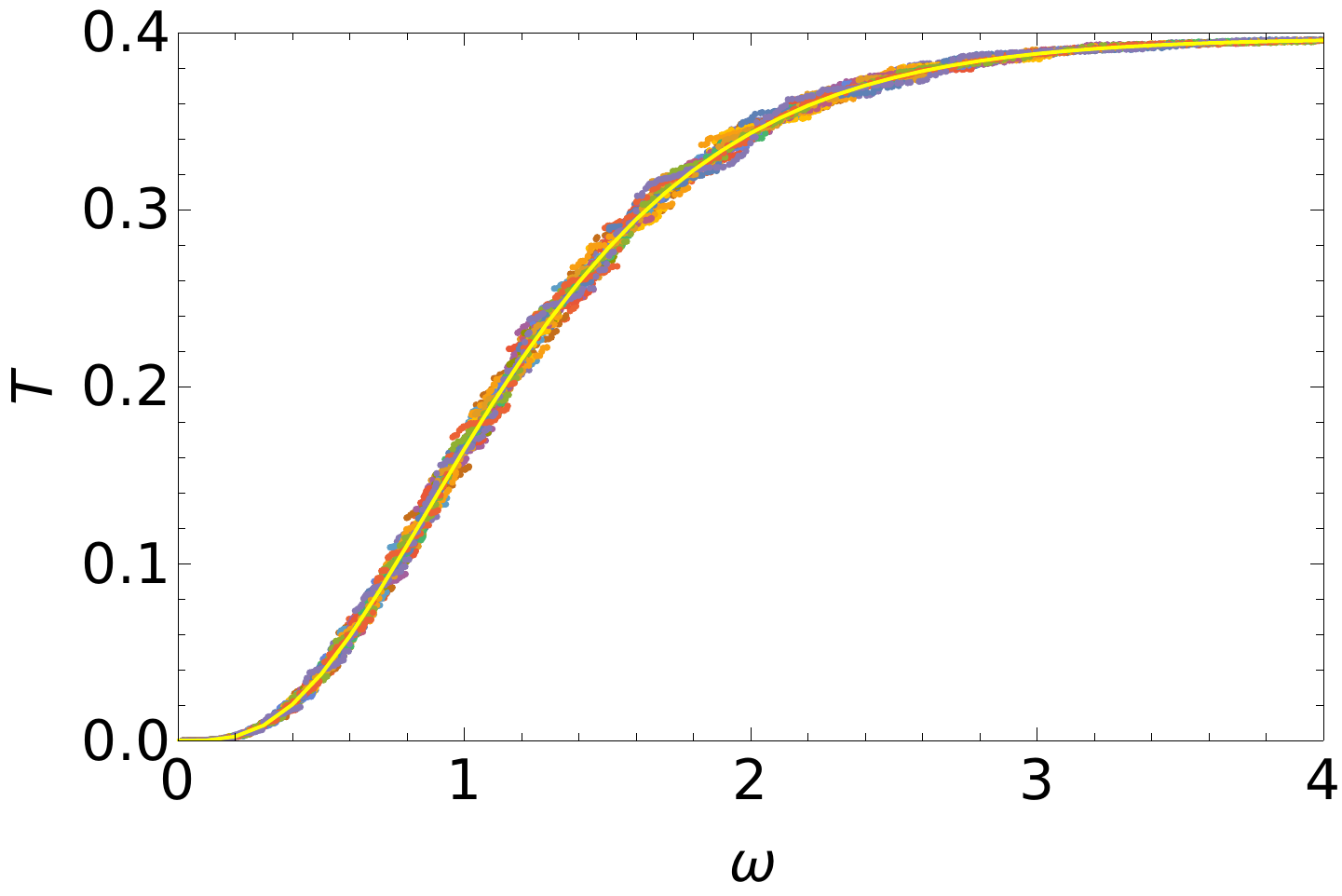}
\caption{Dots: calculation of the integrated response function for a pure three-body continuum for 20 sets of independently chosen Gau{\ss}ian basis functions; the solid yellow line is the analytic result Eq.~\eqref{eq:I0I1}.}
    \label{fig:LIT3b}
\end{figure}

As can be seen in Fig.~\ref{fig:LIT3b}, there is a small but definite scatter in the fully numerical (SVM with a suitably random choice of basis functions) calculations, which agrees very well with the analytical result. Thus we have reached stage one: we are confident that we can deal with a three-body continuum correctly. Now we look at coupled channels.

\section{Two-body bound state and 3 body continuum}\label{sec:full}

We now increase complexity and add to the three-body system a set of scalar (and thus radial) P\"oschl-Teller two-particle potentials; its details are discussed in detail in the Appendix A. In the spirit of our motivation to study the $^3$He/$^3$H systems, two of the particles are assumed to be identical bosons (they could also be fermions with opposite spins), and the third is distinguishable. We assume the potential between identical particles is too weak for a bound state (using $\lambda_{1}=1$), but that the ones between any of the two identical and the third particle is strong enough for a bound state (like in the $^3S_1$ 2N system,
using $\lambda_{2}=\lambda_{3}=2.2$).  Thus
\begin{equation}
	H= -\frac{1}{2} \sum_i \nabla^2_i-
	\sum_{i<j}\frac{\lambda_k(\lambda_k+1)}{2}\frac{1}{\cosh(r_{ij})^p}\,, \label{eq:CC}
\end{equation} 
where $k$ is ``the missing index in $ij$", e.g., $k=3$ for $ij=12$. We use $p=2$ in the following.

We assume the third particle is charged (or the two identical particles are and the third one is neutral, which is equivalent), and look at excitation due to a dipole operator,
relative to the CoM. This  gives rise to an $L=1$ state in the $\eta_3$ coordinate, leaving the other coordinate unaffected.

Further details can be found in the Appendix.

\subsection{Numerical calculation}

\begin{figure}
    \centering
    \includegraphics[width=\columnwidth]{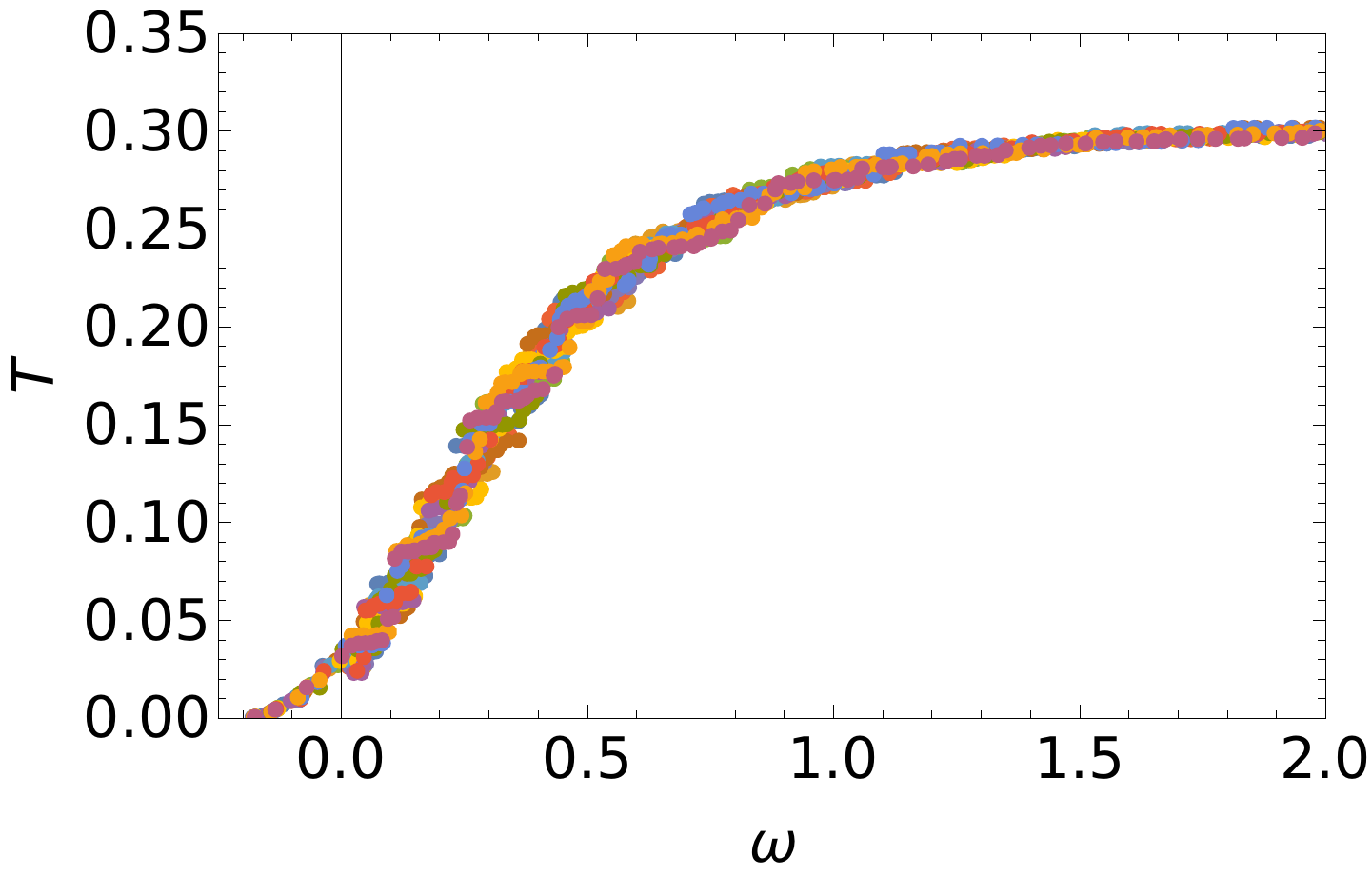}
    \caption{Dots: calculation of the integrated response function  for the coupled-channel problem Eq.~(\ref{eq:CC}). We use
    14 sets of independently chosen Gau{\ss}ian basis functions, indicated by dots of different colour.}
    \label{fig:LITfull}
\end{figure}

As we can see in Fig.~\ref{fig:LITfull}, we get a good description of the integrated response for energies $\omega<0$, but the results spread out directly above the three-particle threshold at $\omega=0$. The scatter of the individual calculations is much larger than in the absence of final state interactions, as shown in Fig.~\ref{fig:LIT3b}. So what is the origin of this? First of all, it is not the effect of channel coupling. While it does play a role, the integrated response distribution shows in the worst case scenario a singularity in the second derivative w.r.t. energy, and thus is barely visible in a plot. The lowest points correspond to cases where some response is lacking directly below zero or directly above zero, i.e., in the pure two-body channel, (typically because there is no state near zero in that specific calculation).

It is not immediate clear from the figure that each individual calculation has 4-to-5 states below zero energy. Actually, a lot of the problematic behaviour is due to states that carry no or very little dipole response. That is not a surprise:  the response at low energy must be driven by the two-body channel (since the phase space of the three-body channel opens very slowly), but there are a large number of different configurations of (almost) zero response states. 

\begin{figure}
    \centering
    \includegraphics[width=\columnwidth]{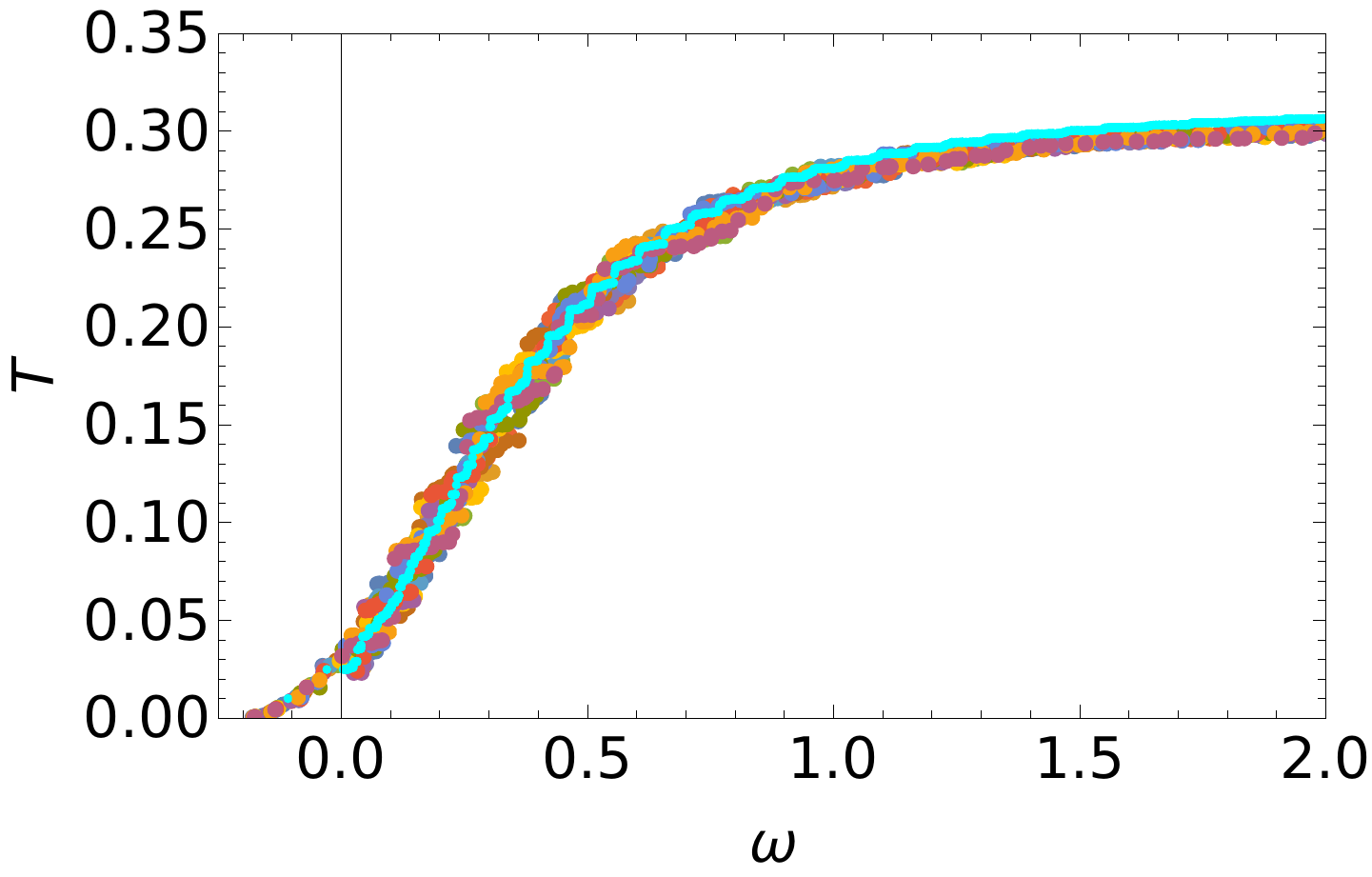}
    \caption{Result from the hyperspherical harmonic calculation, as cyan dots, overlaid on the SVM results, Fig.~\ref{fig:LITfull}.}
    \label{fig:LITcompfull}
\end{figure}

That leads to the question whether this feature is peculiar to the SVM method, or whether more structured computational schemes can resolve this issue. To that end,
we also tackle this model using a hyperspherical harmonics approach in the formalism set out in Ref.~\cite{ballot_application_1980} using a simple harmonic oscillator basis; see Appendix  \ref{app:HH} for details.
For the angular-momentum-zero ground state, we need equal angular momentum on both Jacobi coordinates. For the $J=1$ state caused by the dipole excitation, which only acts on one of the two Jacobi coordinates, we use angular momenta $0\otimes1$ and $2\otimes 1$, $2 \otimes3$, $4 \otimes3$, etc. We use all $L$'s that match the choices made for $J=0$,  with the odd angular momentum one larger than the largest one in the $J=0$ state. The values of $L$ used link to the hyperspherical quantum number $K=2\kappa+L_1+L_2$. Thus we see that, with $\kappa=0$, the cut-off in $K$ is linked to the maximal values of $L$; for example, $J=0$ implies $L_\text{max}=K_\text{max}/2$.
One encounters some difficulties with the hyperspherical approach: while a small $\hbar\omega$ is needed for closely spaced states and a description of the 2+1 channel, it makes it also highly non-trivial to describe the ground state in sufficient detail; for the calculation reported here ($\hbar\omega=0.04$), we use 90 basis states for each value of $K$, and use $K$ up to $16$, and thus $L$ up to 8 for $J=0$ and $9$ for $J=1$). The bound state energy is found as $E=-0.648$, which is slightly larger than the value using the SVM ($E=-0.652$). Since the method provides a strict upper bound, the SVM result therefore is objectively better. Obtaining any sensible values for the two-body bound state as a continuum threshold is even more difficult. Its best estimate from the SVM is $E=-0.172$, with a number of additional states below zero energy, whereas for the HH method we only find two, with the lowest at $E=-0.108$--and even that requires substantial work. Again, the SVM result provides a better bound to the true value. As we can see in Fig.~\ref{fig:LITcompfull}, the strength calculated in the HH calculation is a few percent larger than that in the SVM ones. We have been able to trace this to the slightly poorer choice of 3-body bound state wave function, as discussed above.

Overall we find that there is a great similarity between the two sets of calculations. It is more difficult to find the threshold in the HH calculation--but there exist dedicated methods to improve this \cite{marcucci_hyperspherical_2020}. That also links to the lack of spectral density below $\omega=0$. Interestingly enough, the HH calculation shows the same horizontal jumps as the SVM calculations, but more regularly spaced. Specifically, the HH flat region seen near $\omega=0$  agrees with the average behaviour of the SVM calculations. We thus conclude that for this type of calculation, an SVM method with a suitable basis choice methodology can be highly successful and efficient and makes it probably easier to  extract a smooth result---but at the price of the need to do a number of calculations due to the statistical nature of the process.

\section{Conclusions}\label{sec:conc}

In this paper, we have explored a novel approach to extracting response functions from a bound-state method that uses a basis of square-integrable functions. This provides an alternative to the LIT, and may have advantages over that method in extracting responses of systems of three or more particles.

Our approach uses the SVM to solve the inhomogenous Schr\"odinger equation that gives the response of the wave function to an external field. The stochastically-chosen basis of this method has advantages over the commonly-used hyperspherical harmonics for describing channels with cluster structures, such as appear in break-up of systems of three of more particles, and can describe both thresholds and reponse reasonably well.

Any method using a square-integrable basis generates a response function in the form of a set of $\delta$ distributions. The LIT folds this with a Lorentzian to obtain a continuous function of energy. With care, this can be inverted to yield a continuum response function. In contrast, we consider integrated the response function. This is a stepwise continuous function of energy, with steps at random positions depending on the particular SVM basis. By running a sufficient number of independent SVM calculations, we are able to make a robust fit of a smooth function to the ensemble of integrated response functions. The derivative of this function with respect to energy then provides our approximation to the physical response function. 

We have successfully tested our method using a simple model of two particles interacting via a P\"oschl-Teller potential, where we find excellent agreement with analytic results. We have also applied it to a more realistic two-body system, namely deuteron break-up using the Argonne V18 potential. The agreement with the data is very good, and similar to that of other approaches that consider only the one-body $E1$ operator. We have also used a dispersive method to calculate the real parts of the dynamical polarisabilities corresponding to the deuteron response functions, finding good agreement with older calculations using the LIT.

Our ultimate goal is to apply the method to photodisintegration of and Compton scattering from light nuclei including $^3$H, $^3$He and $^4$He. As a preparation for this, we have applied  it to a simple three-particle model, with an interaction that generates a two-particle bound state. Like the realistic three-body systems, this has two- and three-body continuum channels. Below the three-body threshold, SVM bases give a good description of the strength in the two-body break-up channel, in contrast to HH ones. Above this threshold, the spread of the different bases is wider and so a larger ensemble of them is required for a good fit of a continuous function to the integrated response. The region immediately above this threshold has proved to be a challenging one for both SVM and HH bases, and there are indications that the response there is not well described. We plan to examine this further in future.

One aspect of physics that we have not yet explored in this approach is the treatment of resonant structure in the continuum. This will be needed for applications to $^4$He and heavier nuclei. We hope to address this issue in future work, before applying the method to response functions of light nuclei.
\begin{acknowledgments}
We are indebted to G.~Orlandini and W.~Leidemann for discussions and encouragements which stretch back several years. This work was supported by the UK Science and Technology Funding Council [grant number ST/V001116/1] (all but HWG) and by the US Department of Energy under contract DE-SC0015393 (HWG, JK). Additional funds for HWG were provided by an award of the High Intensity Gamma-Ray Source HI$\gamma$S of the Triangle Universities Nuclear Laboratory TUNL in concert with the Department of Physics of Duke University, and by George Washington University: by the Office of the Vice President for Research and the Dean of the Columbian College of Arts and Sciences; and by an Enhanced Faculty Travel Award of the Columbian College of Arts and Sciences. His research was conducted in part in GW's Campus in the Closet.
\end{acknowledgments}

\bibliography{references}
\newpage
\begin{widetext}
\appendix

\section{SVM approach to Two-Channel Model}\label{app:model}
We consider three inequivalent equal mass particles, with Hamiltonian
\begin{equation}
	H= -\frac{1}{2} \sum_i \nabla^2_i-
	\sum_{i<j}\frac{\lambda_k(\lambda_k+1)}{2}\frac{1}{\cosh(r_{ij})^p}\,, \label{eq:pots}
\end{equation} 
where $k$ is ``the missing index in $ij$", e.g., if $ij=12$, $k=3$.
We choose the same value for $\lambda_k$, $1<\lambda_k<2$ for $k=2,3$ and $0<\lambda_1<1$, so that there is a two-body bound state between the two pairings of ``unlike" particles, but not between the one pair of similar (but distinguishable!) particles.
We shall also assume that only the third particle is charged, and will be interested in dipole excitations relative to the centre of mass.

We shall tackle this problem in terms of the ``T" Jacobi coordinates but will freely exchange between the three equivalent choices, since we shall only use correlated Gau{\ss}ian which can be trivially decomposed in each of the three choices of Jacobi coordinates.

We introduce the Jacobi coordinates, labelled by the third particle index,
\begin{align}
	\vec\xi_3&=\vec{r}_2-\vec{r}_1\,,\nonumber\\
	\vec\eta_3&=(\vec{r}_3-(\vec{r}_2+\vec{r}_1)/2)\,,\label{Eq:Jacobi}\\
	\vec R&=(\vec r_1+\vec r_2+\vec r_3)/3\,,\nonumber
\end{align}
and cyclic permutations.

We will use correlated Gau{\ss}ian wave function of the form
\begin{equation}
\phi(\{\vec r_i\})=
\left(\frac{\det A}{2\pi}\right)^{3/4}
\exp(-\frac{1}{2}A_{ij} \vec r_i \cdot \vec r_j)\,,
\end{equation}
where as usual we add an artificial centre-of-mass Gau{\ss}ian, which we later will remove by taking $a_0\rightarrow 0$, and thus
\begin{equation}
	A =\begin{pmatrix}
		a_0/9-b_2-b_3& a_0/9+b_3 &a_0/9+b_2\\
		a_0/9+b_3&a_0/9-b_1-b_3&a_0/9+b_1\\
		a_0/9+b_2&a_0/9+b_1&a_0/9-b_1-b_2
	\end{pmatrix}\,.
\end{equation}
Using the parametrisation 
\begin{align}
a_0&= \chi,\nonumber\\
b_1&=\gamma-\beta/2, b_2=-\gamma-\beta/2,b_3=-\alpha+\beta/4\,,
\end{align}
or
\begin{align}
\chi&= a_0,\nonumber\\
\alpha &=-\frac{1}{4}(b_1+b_2)- b_3,
\beta = -b_1-b_2,\gamma = (b_1-b_2)/2\,,
\end{align}
we can also write the wave function as
\begin{align}
\phi_{\alpha\beta\gamma;\chi}(\{\vec r_i\})&=\mathcal{N}\exp(-\frac{1}{2}\chi R^2)\nonumber\\
&\times\exp(-\alpha/2 \xi_3^2-\beta/2 \eta_3^2-\gamma \vec \xi_3\cdot \vec \eta_3)\,,\label{eq:wf2}
\end{align}
Calculating the determinant of $A$,  $\det(A)=\chi \left(\alpha  \beta -\gamma ^2\right)$, we see that we must require hat $|\gamma|<\sqrt{\alpha\beta}$.

We shall act with the ``dipole" operator $d_z=z_3-Z=\frac{2}{3} \eta_{3z}$.
We will also want to express this in term of the other coordinates, in other words:
\begin{align}
d_z&=\frac{2}{3} \eta_{3z}\nonumber\\
&=-\frac{1}{3}\eta_{1z}+\frac{1}{2}\xi_{1z}\nonumber\\
&=-\frac{1}{3}\eta_{2z}-\frac{1}{2}\xi_{2z}\,.
\end{align}
\subsection{Ground state}
We first concentrate on the $L=0$ ground state.	
Calculating the kinetic energy is simple:
\begin{equation}
K-K_\text{CoM}=-\nabla^2_{\xi_3}-\frac{3}{4}\nabla^2_{\eta_3}\,.
\end{equation}
When acting between two wave functions of the form \eqref{eq:wf2} we get
\begin{multline}
\bra{\alpha'\beta'\gamma'}K-K_\text{CoM}\ket{\alpha\beta\gamma}=\int \dd[3]\xi_3 \dd[3]\eta_3 \,\phi_{\alpha'\beta'\gamma'}
 \phi_{\alpha\beta\gamma}\\
\left((-\alpha \vec \xi_3 -\gamma\vec \eta_3)\cdot(-\alpha' \vec \xi_3 -\gamma'\vec \eta_3)
 +\frac{3}{4}(-\gamma \vec \xi_3 -\beta\vec \eta_3)\cdot(-\gamma' \vec \xi_3 -\beta'\vec \eta_3)\right) \,.
\end{multline}
So the final integral we need to do is of the form
\begin{multline}
\int \dd[3]\xi_3 \dd[3]\eta_3 \, \left(
\vec\eta_3\cdot  \vec\xi_3  \left(\alpha  \gamma'+\alpha' \gamma +\frac{3 \beta  \gamma'}{4}+\frac{3 \beta' \gamma }{4}\right)+\xi_3 ^2 \left(\alpha  \alpha'+\frac{3 \gamma  \gamma'}{4}\right)+\eta_3 ^2 \left(\frac{3 \beta  \beta'}{4}+\gamma  \gamma'\right)\right)\\
\exp(-\frac{1}{2}(\alpha+\alpha')\xi_3^2-\frac{1}{2}(\beta+\beta')\eta_3^2
-(\gamma+\gamma')\vec \xi_3\cdot \vec \eta_3)\\
=\frac{3}{4}
\frac{ 3 \beta  \beta' (\alpha +\alpha')+4 \alpha  \alpha' (\beta +\beta')-4 \alpha  \gamma'^2-4 \alpha' \gamma ^2-3 \beta  \gamma'^2-3 \beta' \gamma ^2}{(\alpha +\alpha') (\beta +\beta')- (\gamma +\gamma')^2}\braket{\alpha'\beta'\gamma'}{\alpha\beta\gamma}
\,.
\end{multline}
Here 
\begin{equation}
\braket{\alpha'\beta'\gamma'}{\alpha\beta\gamma}=\frac{(\alpha\beta-\gamma^2)^{3/4}(\alpha'\beta'-{\gamma'}^2)^{3/4}}{\left((\alpha+\alpha')(\beta+\beta')-(\gamma+\gamma')^2\right)^{3/2}}.
\end{equation}

If $a_0$ is zero, we can also re-express this in terms of the $b$'s:
\begin{align}
  &\bra{\{b'\}}K-K_\text{CoM}\ket{\{b\}}=
 - \frac{3}{(b_1+b'_1)(b_2+b'_2)+(b_2+b'_2)(b_3+b'_3)+(b_3+b'_3)(b_1+b'_1)}\nonumber\\&\times \biggl( (b_1+b'_1)(b_2+b'_2)(b_3+b'_3)-b_1b_2b_3-b'_1b'_2b'_3+\nonumber\\&\qquad
 b_1b'_1(b_2+b'_2+b_3+b'_3)+b_2b'_2(b_3+b'_3+b_1+b'_1)+b_3b'_3(b_1+b'_1+b_2+b'_2)\biggr)
 \,.
\end{align}
If $b=b'$, we find $-3(b_1+b_2+b_3)/2$.

The potential contributions are a bit more complex, and are best done in terms of the cyclic permutations of $\alpha$, $\beta$ and $\gamma$. Supposing we have those at hand, we can just work through the example of one of these
\begin{multline}
\bra{\alpha'\beta'\gamma'}\frac{1}{\cosh(\xi_3)^p}\ket{\alpha\beta\gamma}=\mathcal{N}\int \dd[3]\xi_3 \exp\left(-\frac{1}{2}\left((\alpha+\alpha') -\frac{(\gamma+\gamma')^2}{\beta+\beta'}\right) \xi_3^2\right)\frac{1}{(\cosh \xi_3)^p}\nonumber \\
\times\int \dd[3]\eta'_3\exp(-\frac{1}{2}(\beta+\beta'){\eta'_3}^2)\nonumber\\
=\braket{\alpha'\beta'\gamma'}{\alpha\beta\gamma} I\left((\alpha+\alpha') -\frac{(\gamma+\gamma')^2}{\beta+\beta'}\right)\nonumber\\
=\braket{\{b'\}}{\{b\}} I_0\left(-\frac{(b_1+b'_1)(b_2+b'_2)+(b_2+b'_2)(b_3+b'_3)+(b_3+b'_3)(b_1+b'_1)}{b_1+b'_1+b_2+b'_2} \right)
\end{multline}
where 
\begin{equation}
I_0(\lambda)= \sqrt{2/\pi} \int r^2\dd r \exp(-\frac{1}{2} r^2) \frac{1}{\cosh (r/\sqrt{\lambda})^p}.
\label{eq:I0}
\end{equation}
We make use of the cyclic permutations of the $b$'s to get an expression for $\alpha$, $\beta$ and $\gamma$ relevant to each of the three potentials,
\begin{align}
\alpha = \frac{1}{4} (-b_1-b_2)- b_3,\beta = -b_1-b_2,\gamma = (b_1-b_2)/2\,.
\end{align}

\subsection{Excited state}
The $L=1,M=0$ excited state we shall consider is again of the correlated Gau{\ss}ian form
\begin{equation}
\phi_1(\{\vec r_i\})=
\mathcal{N}
\left(\sum_i c_i z_i\right)
\exp(-\frac{1}{2}A_{ij} \vec r_i \cdot \vec r_j)\,,
\end{equation}
where $\sum_ic_i=0$.
Here
\begin{equation}
\mathcal{N}=\left(\frac{\det A}{2\pi}\right)^{3/4}
\left((A^{-1})_{ij}c_ic_j\right)^{1/2}\,,
\end{equation}
where we have used
\begin{equation}
\partial_{A_{ij}} \det A=\det A\, (A^{-1})_{ij}.
\end{equation}
We now bring this into a Jacobi form as
\begin{equation}
\braket{\vec \xi,\vec\eta}{\alpha\beta \gamma c}=\mathcal{N}\exp(-\frac{1}{2}\chi R^2)\left(\frac{c_1+c_2-2c_3}3 \eta_{3z}+\frac{c_1-c_2}2 \xi_{3z}\right)\exp(-\alpha/2 \xi_3^2-\beta/2 \eta_3^2-\gamma \vec \xi_3\cdot \vec \eta_3)\,,\label{eq:wf3}
\end{equation}
We can express the norm as
\begin{equation}
	\mathcal N_{\alpha\beta\gamma c}=2^{3/2}\left(\alpha\beta-\gamma^2
		\right)^{5/4}/( c_m^2 \beta+c_p^2 \alpha-2c_mc_p\gamma)^{1/2},
\end{equation}
and the overlap
\begin{equation}
	\braket{\alpha'\beta'\gamma'c'}{\alpha\beta\gamma c}
	=\mathcal N_{\alpha\beta\gamma c}\mathcal N_{\alpha'\beta'\gamma' c'}
	\frac{c_m c'_m (\beta+\beta')+c_p c'_p (\alpha+\alpha')-(c_mc'_p+c'_mc_p)(\gamma+\gamma')}{((\alpha+\alpha')(\beta+\beta')-(\gamma+\gamma')^2)^{5/2}}.
\end{equation}
The kinetic energy takes the form
\begin{multline}
\bra{\alpha'\beta'\gamma'c'}K-K_\text{CoM}\ket{\alpha\beta\gamma c}=\int \dd[3]\xi_3 \dd[3]\eta_3 \,\phi_{\alpha'\beta'\gamma'}
 \phi_{\alpha\beta\gamma}\Biggl(\\
\left((-\alpha \vec \xi_3 -\gamma\vec \eta_3)\cdot(-\alpha' \vec \xi_3 -\gamma'\vec \eta_3)
 +\frac{3}{4}(-\gamma \vec \xi_3 -\beta\vec \eta_3)\cdot(-\gamma' \vec \xi_3 -\beta'\vec \eta_3)\right)\\
 \times \left(-\frac{c'_1+c'_2-2c'_3}3 \eta_{3z}+\frac{c'_1-c'_2}2 \xi_{3z}\right)\left(-\frac{c_1+c_2-2c_3}3 \eta_{3z}+\frac{c_1-c_2}2 \xi_{3z}\right) \\
+ \frac{c'_1+c'_2-2c'_3}3 \frac{c_1+c_2-2c_3}3+\frac{3}{4} \frac{c'_1-c'_2}2\frac{c_1-c_2}2 \Biggr) 
\end{multline}
\begin{align}
=
\Biggl(
c_m c'_m &\biggl(-12 (\beta +\beta') \left(\alpha  \gamma'^2+\alpha' \gamma ^2\right)-8 \gamma  \gamma' (\alpha +\alpha') (\beta +\beta')+12 \alpha  \alpha' (\beta +\beta')^2\nonumber \\&
+3 \beta  \beta' (\alpha +\alpha') (\beta +\beta')-3 \beta' \gamma ^2 (\beta +3 \beta')+12 \beta  \beta' \gamma  \gamma'-3 \beta  \gamma'^2 (3 \beta +\beta')+8 \gamma  \gamma' (\gamma +\gamma')^2\biggr)\nonumber \\
+c_m c'_p& \biggl(3 (\alpha +\alpha') \left(2 \beta ^2 \gamma'-\beta  \beta' (3 \gamma +\gamma')\right)-4 \alpha  \alpha' (\beta +\beta') (\gamma +3 \gamma')\nonumber \\&
+4 (\gamma +\gamma') \left(3 \alpha  \gamma'^2+\alpha' \gamma  (\gamma -2 \gamma')\right)+8 \alpha'^2 \gamma  (\beta +\beta')+3 (\gamma +\gamma') \left(\beta  \gamma' (\gamma'-2 \gamma )+3 \beta' \gamma ^2\right)\biggr)\nonumber \\
+c'_m c_p &\biggl(8 \alpha ^2 \gamma' (\beta +\beta')-4 \alpha  \alpha' (\beta +\beta') (3 \gamma +\gamma')+4 (\gamma +\gamma') \left(\alpha  \gamma' (\gamma'-2 \gamma )+3 \alpha' \gamma ^2\right)\nonumber \\&
-\alpha  \beta' (3 \beta  (\gamma +3 \gamma'))-\alpha  \beta' (-6 \beta' \gamma )+3 \alpha' \beta' (2 \beta' \gamma -\beta  (\gamma +3 \gamma'))+3 (\gamma +\gamma') \left(3 \beta  \gamma'^2+\beta' \gamma  (\gamma -2 \gamma')\right)\biggr)\nonumber \\
+c_p c'_p &\biggl(-12 \left(\alpha ^2 \gamma'^2+\alpha'^2 \gamma ^2\right)-3 (\alpha +\alpha') \left(2 \gamma  \gamma' (\beta +\beta')+3 \beta  \gamma'^2+3 \beta' \gamma ^2\right)\nonumber \\&
+9 (\alpha +\alpha')^2 (\beta  \beta')+4 \alpha  \alpha' (\alpha +\alpha') (\beta +\beta')-4 \alpha  \alpha' \left(\gamma ^2-4 \gamma  \gamma'+\gamma'^2\right)+6 \gamma  \gamma' (\gamma +\gamma')^2\biggr)
\Biggr)\nonumber \\
&\times\frac{1}{4 \left((\alpha+\alpha')  (\beta +\beta')-(\gamma +\gamma')^2\right)^2}\braket{\alpha'\beta'\gamma'}{\alpha\beta\gamma}\,.
\end{align}
Here 
\begin{align}
c_m&=(c_2-c_1)/2\,,\\
c_p&=-(c_1+c_2-2c_3)/3=c_3\,.
\end{align}
The potential can also be evaluated:
\begin{multline}
\bra{\alpha'\beta'\gamma'c'}\frac{1}{\cosh(\xi_3)^p}\ket{\alpha\beta\gamma c}=\mathcal{N}\mathcal{N'}\int \dd[3]\xi_3 
 (c_p c'_p +\frac{(c_m(\beta+\beta')-c_p(\gamma+\gamma'))(c'_m(\beta+\beta')-c'_p(\gamma+\gamma')}{\beta+\beta'}\xi_{3z}^2)\\ \times
\exp\left(-\frac{1}{2}\left((\alpha+\alpha') -\frac{(\gamma+\gamma')^2}{\beta+\beta'}\right) \xi_3^2\right)\frac{1}{\cosh^p \xi_3}\nonumber \\
\times \frac{(2\pi)^{3/2}}{(\beta+\beta')^{5/2}}\nonumber\\
=\braket{\alpha'\beta'\gamma'c'}{\alpha\beta\gamma c} V_2\left(c_pc'_p,\frac{(c_m(\beta+\beta')-c_p(\gamma+\gamma'))(c'_m(\beta+\beta')-c'_p(\gamma+\gamma')}{\beta+\beta'},(\alpha+\alpha') -\frac{(\gamma+\gamma')^2}{\beta+\beta'}\right)\,.
\end{multline}
Here $V_2$ is a short-hand for the integral
\begin{align}
V_2(c,\epsilon,\delta)&=\frac{\int \dd[3]\xi_3 \exp(-\delta \xi_3^2/2)(c +\epsilon\xi_{3z}^2)\frac{1}{\cosh^p \xi_3}}{\int \dd[3]\xi_3 \exp(-\delta \xi_3^2/2)(c +\epsilon\xi_{3z}^2)}\nonumber\\
&=\frac{\int r^2 \dd r \exp(-\delta r^2/2)(c +\frac{1}{3}\epsilon r^2)\frac{1}{\cosh^p r}}{\int r^2 \dd r \exp(-\delta r^2/2)(c +\frac{1}{3}\epsilon r^2)}.
\end{align}
The integral in the denominator is $\sqrt{\pi/2}(c\delta +\epsilon)/\delta^{5/2}$. The one in the numerator is best dealt with using the asymptotic expansion
\begin{align}
&\int r^2 \dd r \exp(-\delta r^2/2)(c +\frac{1}{3}\epsilon r^2)\frac{1}{\cosh^p r}\nonumber\\
&=\delta^{-3/2}\int r^2 \dd r \exp(- r^2/2)\left(c +\frac{1}{3}\frac{\epsilon}{\delta} r^2\right)\frac{1}{\cosh^p (r\delta^{-1/2})}\\
&=c\delta^{-3/2}\sqrt{\pi/2} I(\delta)+\frac{1}{3}\epsilon3\delta^{-5/2}\sqrt{ \pi/2}I_2(\delta)
\,,
\end{align}
where
\begin{align}
	I_2(\delta)&=\frac{1}{3}\sqrt{2/\pi}
	\int r^4\dd r \exp(-\frac{1}{2} r^2) \frac{1}{\cosh^p (r/\sqrt{\delta})},
\end{align}
and $I_0$ is specified in q.~\eqref{eq:I0}.
The numerical parametrisation can be tackled through an asymptotic expansion in $\delta$, which is then turned into a high-order $[n/n+1]$ Pad\'{e} approximant in $1/\delta$. The difference in powers in the Pad\'{e} gives the correct zero for $\delta=0$.

So, finally, 
\begin{align}
V_2(c,\epsilon,\delta)=\frac{c \delta I_0(\delta)+\epsilon I_2(\delta)}{c \delta+\epsilon}\,.
\end{align}

\section{Hyperspherical Harmonics}\label{app:HH}
We also tackle the problem of Appendix~\ref{app:model} using a hyperspherical harmonics approach; we follow the formalism as set out in Ref.~\cite{ballot_application_1980}, using a simple harmonic oscillator basis.

\subsection{Basis used}
Following the standard conventions we label the Jacobi coordinates as $\xi_1$ and $\xi_2$, which equal $\xi_3$ and $\sqrt\frac{4}{3}\eta_3$ as used above, and thus
\begin{align}
\vec \xi _1&=\vec r_1-\vec r_2,\nonumber\\
\vec \xi _2&=\sqrt{\frac{4}{3}} \left(\vec r_3-\frac{\vec r_1+\vec r_2}{2}\right),\nonumber\\
\vec R_\text{cm}&=\frac{1}{3} (\vec r_1+\vec r_2+\vec r_3),
\end{align}
and then define
\begin{equation}
\xi_1=\xi\sin\phi,\quad \xi_2=\xi\cos\phi\,.
\end{equation}
we find that
\begin{equation}
\nabla^2_{\xi_1} +\nabla^2_{\xi_2}=
\frac{\partial ^2}{\partial^2 \xi }+\frac{5}{\xi } \frac{\partial }{\partial \xi }+\frac{1}{\xi ^2}\left(\frac{\partial ^2}{\partial^2 \phi}+4 \cot (2 \phi ) \frac{\partial }{\partial \phi }-  \left(
\csc ^2(\phi )L_1^2+ \sec ^2(\phi )L_2^2\right)\right)\,.
\end{equation}
The integration volume is
\begin{equation}
\dd^3\xi_1d^3\xi_2=\xi ^5 \dd\xi  \sin ^2(\phi ) \cos ^2(\phi )\dd\phi \,\dd\Omega_1\dd\Omega_2\,.
\end{equation}
We now define the 6-dimensional harmonic oscillator state as the eigenvalues of 
\begin{equation}
H=-\frac{1}{2}(\nabla^2_{\xi_1} +\nabla^2_{\xi_2})+\frac{1}{2}\omega^2 \xi^2\,.
\end{equation}
As usual we separate in a radial and (hyper-)angular part. It is easy to show that the latter part has the normalised eigenfunctions
\begin{equation}
\varphi_{\kappa,L_1,M_1,L_2,M_2}(\phi,\Omega_1,\Omega_2)=\mathcal{N}_{\kappa  L_1 L_2}\sin^{L_1+1/2}(\phi)\cos^{L_2+1/2}(\phi) P^{(L_2+1/2,L_2+1/2)}_\kappa(\cos 2 \phi) Y^{L_1}_{M_1}(\Omega_1)Y^{L_2}_{M_2}(\Omega_2)\,.
\end{equation}
These are eigenfunctions of the hyper-angular momentum operator 
\begin{equation}
{\hat K}^2=-\frac{\partial ^2}{\partial^2 \phi}-4 \cot (2 \phi ) \frac{\partial }{\partial \phi }+  
\csc ^2(\phi )L_1^2+ \sec ^2(\phi )L_2^2\,
\end{equation}
 with eigenvalue of $K(K+4)$, where $K=2\kappa+L_1+L_2$. 
 
 The radial wavefunction is now given by Laguerre polynomials,
\begin{equation}
\psi_{Kn}(\xi)=\mathfrak{N}_{K n}e^{- \xi ^2 \omega/2 } \left(\xi  \sqrt{\omega }\right)^K \sqrt{2 \omega ^3 } L_n^{K+2}\left(\xi ^2 \omega \right)\,.
\end{equation}
Here we use normalisation constants
\begin{align}
\mathcal{N}_{\kappa  L_1 L_2}&=\sqrt{\frac{2 \Gamma (\kappa +1) \left(2 \kappa +L_1+L_2+2\right) \Gamma \left(\kappa +L_1+L_2+2\right)}{\Gamma \left(\kappa +L_1+\frac{3}{2}\right) \Gamma \left(\kappa +L_2+\frac{3}{2}\right)}},\nonumber\\
\mathfrak{N}_{K n} &
=
\sqrt{\frac{n!}{(K+n+2)!}}\,.
\end{align}
\subsection{Kinetic energy matrix elements}
As for all harmonic oscillator problems, the matrix elements of the kinetic energy operator are very simple. Since this is a hyper-scalar operator, $K$ and consequently the hyper-radial wave function, are unchanged by the action of the kinetic energy, and we find that
\begin{align}
\left[-\frac{\partial ^2}{\partial^2 \xi }-\frac{5}{\xi } \frac{\partial f}{\partial \xi }+\frac{K(K+4)}{\xi^2}\right]\psi_{Kn}(\xi)=&
\omega(3 + K + 2 n)\psi_{Kn}(\xi) \nonumber\\&+\omega\sqrt{(n+1) (K+n+3)}\psi_{Kn+1}(\xi)\nonumber\\&+\omega \sqrt{n (K+n+2)}\psi_{Kn-1}(\xi)\,.
\end{align}
\subsection{Potential energy matrix elements}
We express the potential energy in a multipole expansion,
\begin{equation}
V(\xi,\phi,\theta_{12})=\sum _{L=0}^{\infty } P_L\left(\cos \left(\theta _{12}\right)\right) V_L(\xi ,\phi )\,.
\end{equation}
We define (the is is the only place where hat denotes a unit vector)
\begin{equation}
z=\cos \left(\theta _{12}\right)=\hat{\xi }_1\cdot \hat{\xi }_2\,,
\end{equation}
and thus
\begin{equation}
V_L(\xi ,\phi )=\frac{2 L+1}{2}  \int_{-1}^1 P_L(z) V(\xi ,\phi ,z) \, dz\,
.
\end{equation}
We assume that the potential energy is a sum of central two-body forces
\begin{equation}
V=\sum_{i<j}V_{ij}\left(r_{ij}\right),
\end{equation}
where we take the $12$ interaction different from the $13$ and $23$ ones, which are assumed to  be equal. After some coordinate algebra
we then find that
\begin{equation}
V(\xi ,\phi ,z)=V_{12}(\xi  \sin\phi )+V_{13}\left(\frac{1}{2} \xi  \sqrt{\sqrt{3} z \sin (2 \phi )+\cos (2 \phi )+2}\right)+V_{13}\left(\frac{1}{2} \xi  \sqrt{-\sqrt{3} z \sin (2 \phi )+\cos (2 \phi )+2}\right)\,.
\end{equation}
The first term is independent of $z$.so only contributes an $L=0$ multipole; in the next two terms  the odd powers in the multipole expansion cancel, and thus $V$ only contains even powers of $L$. If we restrict the calculation to even values we can pick one of the two last forms, and write
\begin{equation}
V(\xi ,\phi ,z)=V_{12} (\xi  \sin\phi )+2 V_{13} \left(\frac{1}{2}\xi  \sqrt{-\sqrt{3} z ( \sin(2 \phi ) )+ \cos(2 \phi ) +2}\right)\,.
\end{equation}
We can now calculate the matrix elements of $V$ between three-body HH states in two parts:
\begin{equation}
\left\langle n K L_1 L_2 J M =0|V|n K' L_1' L_2' J M=0\right\rangle =\left\langle n K L_1 L_2 \left|V_L\right|n' K' L_1' L_2' n'\right\rangle  \left\langle L_1 L_2 JM\left|P_L\left( \cos\theta _{12} \right)\right| L_1' L_2'J M\right\rangle
.
\end{equation}
Here, using the standard notation $\hat L=\sqrt{2L+1}$,
\begin{equation}
\left\langle  L_1 L_2 J M\left|P_L\left( \cos \theta _{12} \right)\right| L_1' L_2'J M\right\rangle = -\left\{
\begin{array}{lll}
 L_1' &L_2'& J \\L_2&L_1&L\end{array}
\right\}
\widehat{L_2}\widehat{L_1}\widehat{L_2'}\widehat{ L_1'}
\begin{pmatrix}
 L_2' & L & L_2 \\
 0 & 0 & 0 \\
\end{pmatrix}
\begin{pmatrix}
 L_1' & L & L_1 \\
 0 & 0 & 0 
\end{pmatrix}\label{eq:angm}\,,
\end{equation}
see the more detailed explanation in the next section.
The radial matrix element is
\begin{align}
\left\langle n K L_1 L_2\left|V_L\right|n' K' L_1' L_2'\right\rangle =&\mathfrak{N}_{K n}\mathfrak{N}_{K' n'}\int \xi ^5d\xi \,\left(\xi  \sqrt{\omega }\right)^{K'+K}\exp  \left(- \xi ^2 \omega \right)L_n^{K+2}\left(\xi ^2 \omega \right)L_{n'}^{K'+2}\left(\xi ^2 \omega \right)\nonumber\\
&\quad\mathcal{N}_{\kappa  L_1 L_2}\mathcal{N}_{\kappa ' L_1' L_2'}\int d\phi \,V_L(\xi ,\phi ) \sin ^{L_1'+L_1+2}\phi  \cos ^{L_2'+L_2+2}\phi\nonumber\\
&\qquad\qquad P_{\kappa }^{\left(L_1+\frac{1}{2},L_2+\frac{1}{2}\right)}(\cos (2 \phi ) )P_{\kappa '}^{\left(L_1'+\frac{1}{2},L_2'+\frac{1}{2}\right)}( \cos(2 \phi ) )\,,
\end{align}
with
$
K=2 \kappa +L_1+L_2$.

In practical calculations, we  can calculate a set of  $\phi$ integrals once. We can then vary the radial basis size independently, keeping the angular dependence fixed. The $\xi$ integrals are easily calculated by a 200th order Gau{\ss}-Laguerre integration in extended precision.
We also need to evaluate the transition matrix elements.
Using the fact that the ``E1" operator we use can be expressed as $\frac{1}{\sqrt{3}}\xi_{2,z}=\frac{1}{\sqrt{3}}\xi\cos\phi \cos \theta_2$, we calculate
\begin{align}
&\frac{1}{\sqrt{3}}\xi_{2,z}\left\langle n K L_1 L_2 J =1\left|\xi \cos\phi  \cos \theta _2 \right|n' K' L_1' L_2' J=0\right\rangle \nonumber\\
&\quad =\mathfrak{N}_{K n}\mathfrak{N}_{K' n'}\frac{1}{\sqrt{3\omega }}\int \xi ^5 d\xi  \exp  \left(- \xi ^2 \omega \right) \left(\xi  \sqrt{\omega }\right)^{K'+K+1}  L_n^{K+2}\left(\xi ^2 \omega \right) L_{n'}^{K'+2}\left(\xi ^2 \omega \right)\nonumber\\
&\quad\times\mathcal{N}_{\kappa ' L_1' L_2'}\mathcal{N}_{\kappa  L_1 L_2} \int d\phi  \sin ^2\left(\phi  \right)   \cos ^3\left(\phi\right)   \sin ^{L_1'+L_1}\left(\phi\right)  \cos ^{L_2'+L_2}\left(\phi \right) P_{\kappa }^{\left(L_1+\frac{1}{2},L_2+\frac{1}{2}\right)}(\cos  (2 \phi )) P_{\kappa '}^{\left(L_1'+\frac{1}{2},L_2'+\frac{1}{2}\right)}(\cos  (2 \phi ))\nonumber\\
&\quad\times \left\langle L_1 L_2 J=1 J_{z}=0\left|\cos\theta _2  \right| L_1' L_2'J=0\right\rangle
\end{align}
and the norm
\begin{align}
&\frac{1}{3}\left\langle n K L_1 L_2 J =0\left|(\xi \cos\phi  \cos \theta _2)^2 \right|n' K' L_1' L_2' J=0\right\rangle \nonumber\\
&\quad =\mathfrak{N}_{K n}\mathfrak{N}_{K' n'}\frac{1}{3 \omega }\int \xi ^5 d\xi  \exp  \left(- \xi ^2 \omega \right) \left(\xi  \sqrt{\omega }\right)^{K'+K+2}  L_n^{K+2}\left(\xi ^2 \omega \right) L_{n'}^{K'+2}\left(\xi ^2 \omega \right)\nonumber\\
&\quad\times\mathcal{N}_{\kappa ' L_1' L_2'}\mathcal{N}_{\kappa  L_1 L_2} \int d\phi  \sin ^2\left(\phi  \right)   \cos ^4\left(\phi\right)   \sin ^{L_1'+L_1}\left(\phi\right)  \cos ^{L_2'+L_2}\left(\phi \right) P_{\kappa }^{\left(L_1+\frac{1}{2},L_2+\frac{1}{2}\right)}(\cos  (2 \phi )) P_{\kappa '}^{\left(L_1'+\frac{1}{2},L_2'+\frac{1}{2}\right)}(\cos  (2 \phi ))\nonumber\\
&\quad\times \left\langle L_1 L_2 J=0\left|\cos^2\theta _2  \right| L_1' L_2'J=0\right\rangle
\end{align}
That means there are two more angular integrals to calculate--the radial ones ae dealt with through the same Gau{\ss}-Laguerre technique.

\subsection{Angular momentum algebra}
The angular momentum part of the matrix element requires some trivial recoupling algebra.
We first look at the matrix element
\begin{equation}
\braOket{L_1'L_2'JJ_z=0}{P_L(\cos\theta_{12})}{L_1L_2JJ_z=0}.
\end{equation}
We write 
\begin{equation}
P_L(\cos\theta_{12})=(-1)^L \frac{4 \pi }{\hat{L}} 
\left[ Y_L\left(\Omega_1\right)\otimes  Y_L\left(\Omega_2\right)
\right]^0
\end{equation}
and use  the reduced matrix element,
\begin{equation}
\braOket{L_1'L_2'JJ_z=0}{ P_L(\cos\theta_{12}) }{L_1L_2JJ_z=0}=\frac{1}{\hat{J}}\braOketred{L_1'L_2'J}{P_L(\cos\theta_{12})}{L_1L_2J},
\end{equation}
as shown in Fig.~\ref{fig:general}. As usual, we re-express the conjugate state in terms of a time reversed one,
$\tilde{Y}^L_M=(-1)^{L+M}\left(Y^{L}_{-M}\right)^*$.
\begin{figure}
\begin{center}\includegraphics[width=0.5\textwidth]{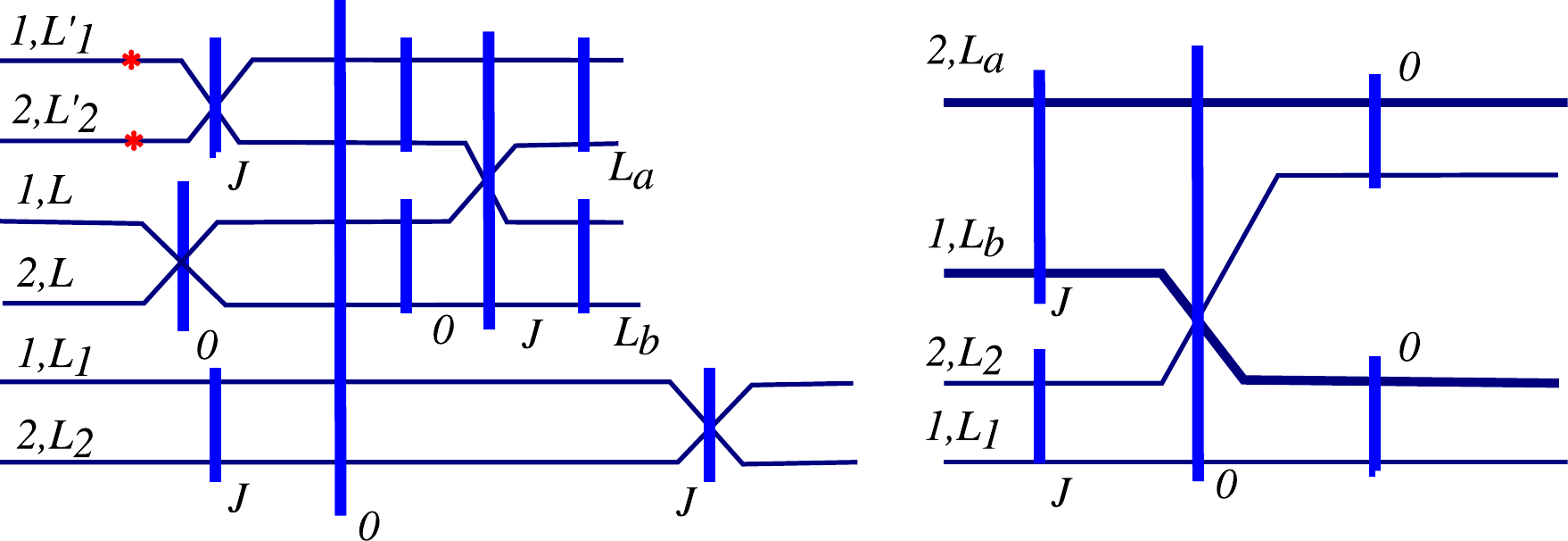}\end{center}
\caption{Reduced matrix element for a general state. A red star denotes a time-reversed state. See \cite{normand_lie_1980} for an explanation of the notation used.}\label{fig:general}
\end{figure}

We can now evaluate the matrix element as in Fig.~\ref{fig:general}. Below, the square brackets denote a square $9J$ symbol, also called a recoupling symbol that arises in the recoupling of 4 angular momenta, \cite{normand_lie_1980}
\begin{align}
&\frac{1}{\hat{J}}(-1)^L\frac{4 \pi }{\hat L}(-1)^{-J+L_1'+L_2'}(-1)^{2 L}
\begin{bmatrix}
 L_2' & L_1' & J \\
 L & L & 0 \\
 L_a & L_b & J 
\end{bmatrix}
(-1)^{-J+L_1+L_2}
\begin{bmatrix}
 L_a & L_b & J \\
 L_2 & L_1 & J \\
 0 & 0 & 0 
\end{bmatrix}\nonumber\\
&\quad\times
(-1)^{L_2'}\frac{\hat{L}\hat{L_2}\hat{ L_2'}}{\sqrt{4 \pi }}
\begin{pmatrix}
 L_2' & L & L_2 \\
 0 & 0 & 0 
 \end{pmatrix}
 (-1)^{L_1'}\frac{\hat{L}\hat{L_1}\hat{L_1'}}{\sqrt{4 \pi }}
\begin{pmatrix}
 L_1' & L & L_1 \\
 0 & 0 & 0 \\
\end{pmatrix},
\end{align}
which results in the expression  in Eq.~\eqref{eq:angm}.

\begin{figure}
\begin{center}\includegraphics[scale=0.5]{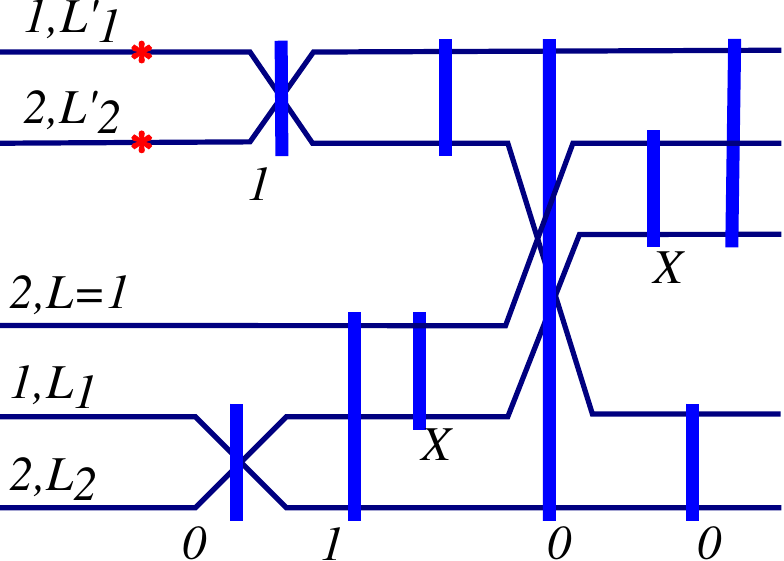}\qquad\includegraphics[scale=0.5]{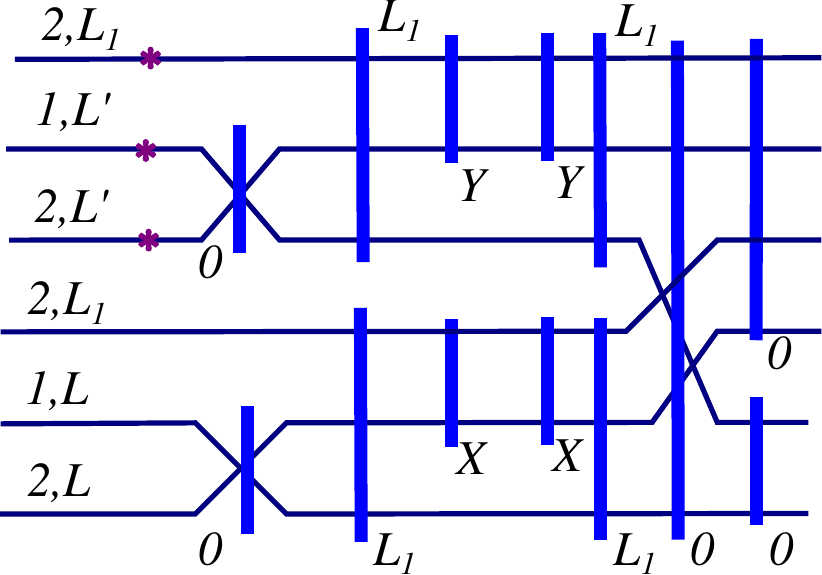}\end{center}
\caption{Reduced matrix element for the overlap (left) and the norm (right) of a state multiplied with $\xi_2$. A red star denotes a time-reversed state.}\label{fig:ovlnrm}
\end{figure}

The overlap and norm are simpler, see Fig.~\ref{fig:ovlnrm}, and we find, expressing $\cos\theta_2$ as $\sqrt{4\pi/3} Y^1_0(\Omega_2)$
\begin{align}
&\braOket{L_1'L_2'J=1J_z=0}{ \cos\theta_2 }{L_1L_2J=0}\nonumber\\
&=\frac{1}{\sqrt{3}}
\sqrt{\frac{4 \pi }{3}} 
(-1)^{L_1'+L_2'-1} \delta _{L_1,L_2}
\begin{bmatrix}
 1 & 0 & 1 \\
L_2& L_1 & 0 \\
 X & L_1 & 1 \\
\end{bmatrix}
\begin{bmatrix}
 L_2' & L_1' & 1 \\
 X & L_1 & 1 \\
 0 & 0 & 0 \\
\end{bmatrix}
(-1)^{L_2'}\frac{\hat 1 \widehat{ L'_2} \widehat{L_2}}{\sqrt{4 \pi }}
\begin{pmatrix}
 L_2' & 1 & L_2 \\
 0 & 0 & 0 \\
\end{pmatrix}
\widehat{L_1}
\nonumber\\&
=(-1)^{L_2'}\frac{\widehat{L_2'}}{\hat 1}
\begin{pmatrix}
 L_2' & 1 & L_2 \\
 0 & 0 & 0 \\
\end{pmatrix},
\end{align}
and (where $L_1=1$ in the figure)
\begin{align}
&\braOket{L'L'J=0}{ \cos^2\theta_2 }{LLJ=0}
\nonumber\\&
=\frac{1}{\widehat{1}}\frac{4 \pi }{3}
\sum _{X,Y}
\begin{bmatrix}
 1 & 0 & 1 \\
 L' & L' & 0 \\
 Y & L' & 1 \\
\end{bmatrix}
\begin{bmatrix}
 1 & 0 & 1 \\
 L & L & 0 \\
 X & L & 1 \\
\end{bmatrix}
\begin{bmatrix}
 Y & L' & 1 \\
 X & L & 1 \\
 0 & 0 & 0 \\
\end{bmatrix}
\frac{1}{4 \pi }\widehat{L}^2 3\widehat{X}
\begin{pmatrix}
 L & 1 & X \\
 0 & 0 & 0 \\
\end{pmatrix}^2
\hat{L}
\nonumber\\&
=\frac{1}{3}\delta _{LL'}\sum _{X=L\pm 1}\hat{X}^2
\begin{pmatrix}
 L & 1& X \\
 0 & 0 & 0 \\
\end{pmatrix}=\frac{\delta _{LL'}}{3}\,.
\end{align}

\end{widetext}
\end{document}